\documentclass[%
 reprint,
 amsmath,amssymb,
 iop,jpcm,
]{revtex4-1}

\usepackage{graphicx}% Include figure files
\usepackage{dcolumn}% Align table columns on decimal point
\usepackage{bm}% bold math
\usepackage{color}

\begin{document}

\title{Phase-field modelling of the effect of density change on solidification revisited:\\
Model development and analytical solutions for single component materials}%ith Force line breaks with \\
% Nonlinear hydrodynamic theory of multicomponent fluid flows

\author{Gyula I. T\'oth}
\affiliation{Interdisciplinary Centre for Mathematical Modelling and Department of Mathematical Sciences, Loughborough University, Epinal Way, Loughborough, LE11 3TU, United Kingdom}
\email{g.i.toth@lboro.ac.uk}

\author{Wenyue Ma}
\affiliation{{School of Mathematics, University of Edinburgh, Peter Guthrie Tait Road, Edinburgh, EH9 3FD, United Kingdom}}


\date{\today}% It is always \today, today,

\begin{abstract}
In this paper the development of a physically consistent phase-field theory of solidification shrinkage is presented. The coarse-grained hydrodynamic equations are derived directly from the $N$-body Hamiltonian equations in the framework of statistical physics, while the constitutive relations are developed in the framework of the standard Phase-field Theory, by following the variational formalism and the principles of non-equilibrium thermodynamics. To enhance the numerical practicality of the model, quasi-incompressible hydrodynamic equations are derived{, where sound waves are absent (but density change is still possible), and therefore the time scale of solidification is accessible in numerical simulations}. The model development is followed by a comprehensive mathematical analysis of the equilibrium and propagating 1-dimensional solid-liquid interfaces for different density-phase couplings. It is shown, that the fluid flow decelerates/accelerates the solidification front in case of shrinkage/expansion of the solid compared to the case when no density contrast is present {between the phases.} Furthermore, {such} a free energy construction is proposed{,} in which {the} equilibrium planar {phase-field} interface is independent from the density-phase {coupling}, and {the equilibrium} interface represents an exact propagating planar interface solution of the quasi-incompressible hydrodynamic equations. Our results are in excellent agreement with previous theoretical predictions.
\end{abstract}
   
%\pacs{}% PACS, the Physics and Astronomy
\maketitle

%\tableofcontents

\section{Introduction}

Microscopic structural change upon solidification of an under-cooled liquid is usually accompanied by a density difference between the initial liquid and the forming solid, which ultimately generate fluid flow. In a system where the solid is more dense than the liquid, extra material is needed to build the solid structure, and therefore the liquid flows towards the solid. In contrast, in water or silicon, for instance, where the material expands upon freezing, the excess mass must be transported away from the front, meaning that the liquid flows away from the solid. The relative density gap is typically around $5\%-10\%$ in simple systems (such as the hard-sphere and the Lennard-Jones model system, water, and simple metals), which suggests that the effect of the density contrast on the morphology of the forming solid pattern might be significant. From the analysis of experimental results for thermal dendrite growth in single compoent systems \cite{PhysRevLett.73.573} it indeed became clear \cite{PINES1996798}, that Stefan's flow (i.e., heat transport by fluid flow) had to be icluded in Ivantsov's original theory \cite{Ivantsov} to elimiate the discrepancy between the experimental data and the theoretical predictions \cite{MCFADDEN1986507}. The tension field created by nuclei in the presence of density difference between the solid and the liquid was first investigated theoretically by Horvay \cite{HORVAY1965195}, while the effect of solidification shrinkage on the speed of a planar solidification front was first studied by Tien and Koump \cite{Tien1970}. Tien and Koump found that fluid flow impeded the solidification front in the presence of shrinkage in binary alloys, and they argued that the solidification front slows down compared to the gapless case, because the extra material transported to the interface region by the fluid flow also must undergo the structural transition. Two decades later Oxtoby and Harrowell developed {the first continuum theory} \cite{doi:10.1063/1.462864} to study the problem. {Their results suggets, that the inverse effect is also present, namely, in case of expansion (the solid/liquid density ratio is less than 1) the solidification front is faster compared to the gapless case. This result nicely accords with the argumentation of Tien and Kuomp: If the extra material transported by the flow to the interface impedes the structural transition, the opposite is expected in case of expansion, when mass is transported away from the interface, and therefore less liquid needs undergo the structural transition. This result is further supported in numerical simulations by Conti \cite{PhysRevE.64.051601}, who studied the effect of fluid flow on the early stage of dendriticl solidification in binary \cite{PhysRevE.67.026117} and monatomic \cite{PhysRevE.69.022601} systems. In order to have access to the time scale of dendrite formation in the numerical simulations, Conti needed to reduce the speed of sound in the system, which was done by using unphysically low bulk moduli. Consequently, his hypothetical material behaved as air from the viewpoint of fluid flow. Our aim here is to fix this problem by \textit{eliminating} sound waves from a hydrodynamical model, together with keeping the possibility of density variations.} The structure of the paper is as follows. In Section II we develop a general theoretical framework of solidification shrinkage for single component materials. The development of the model starts with defining the spatio-temporal solid and liquid mass densities on the level of particles, for which microscopic continuum equations will be derived. This is followed by the derivation of the coarse-grained continuum mechanical equations in the framework of statistical physics. At the end of Section II.A.1, a thermodynamically consistent formulation of the coarse-grained reversible stress tensor is given. In section II.A.2, a thermodynamically consistent constitutive relation for the phase change rate will be given, which will then be followed by a variable transformation providing an \textit{exact} definition of the phase-field. A simple free energy functional containing phase-density coupling will then be defined on the basis of the standard Phase-Field Theory. We will show that the fundamental time scale of the compressible hydrodynamic equations is the time scale of sound waves, which makes the time scale of solidification numerically inaccessible for small thermodynamic driving forces. {Starting from the idea of Lowengrub and Truskinowsky \cite{Lowengrub2617}, we generalise the concept of \textit{quasi-incompressibility} to eliminate sound waves together with keeping variable density, and derive hydrodynamic equations in Section II.A.4}. The derivation of the general framework is followed by the investigation of the planar interface solutions in section II.B. The quasi-incompressible description requires a density-phase relationship, which can be determined by solving the Euler-Lagrange equations for the equilibrium planar solid-liquid interface. In Section II.B.2 we provide the general methodology of finding propagating planar interfaces in both the compressible and quasi-incompressible hydrodynamic framework. In Section III we will apply the model for particular density-phase couplings in the free energy functional. First we investigate the gapless model against homogeneous equilibrium phases, the equilibrium planar solid-liquid interface, propagating 1-dimensional solid-liquid interfaces, and the stability of the bulk phases. This is followed by the same calculations for 2 different density-phase couplings. {The summary and} our concluding remarks are written in Section IV.

\section{Model development}

\subsection{Hydrodynamic phase-field theory of solidification}

\subsubsection{General coarse-grained equations}

Consider a system of $N$ identical particles. Let $\mathbf{\Gamma}(t):=\{\mathbf{r}_1(t),\mathbf{r}_2(t),\dots,\mathbf{r}_N(t),\mathbf{p}_1(t),\mathbf{r}_2(t),\dots,\mathbf{p}_N(t)\}$ denote the trajectory of the system in the $6N$-dimensional phase space, where $\mathbf{r}_i(t)$ and $\mathbf{p}_i(t)$ are the position and momentum of particle $i=1\dots N$, respectively. Let $g_i(t) \in \mathbb{R}$ a{n averaged} bond-order parameter \cite{PhysRevB.28.784,doi:10.1063/1.2977970} quantifying the similarity of the neighbourhood of particle $i$ compared to the perfect crystal structure at time $t$. The microscopic mass density of the solid can be then defined as: $\hat{\rho}_s(\mathbf{r},t) := m\sum_i \phi_i(t) \delta[\mathbf{r}-\mathbf{r}_i(t)]$, where $m$ is the particle mass, $\phi_i(t):=||g_i(t)|| \in [0,1]$ the normalized bond-order parameter, and $\delta(\mathbf{r})=\delta(x)\delta(y)\delta(z)$ the 3-dimensional Dirac-delta function. Analogously, the microscopic mass density operator of the liquid reads: $\hat{\rho}_l(\mathbf{r},t) := m\sum_i [1-\phi_i(t)] \delta[\mathbf{r}-\mathbf{r}_i(t)]$, while the microscopic momentum densities of the phases are given by $\hat{\mathbf{g}}_s(\mathbf{r},t):=\sum_i \phi_i(t) \mathbf{p}_i(t) \delta[\mathbf{r}-\mathbf{r}_i(t)]$ and $\hat{\mathbf{g}}_l(\mathbf{r},t):=\sum_i [1-\phi_i(t)] \mathbf{p}_i(t) \delta[\mathbf{r}-\mathbf{r}_i(t)]$, respectively. The above definitions indicate the following identities:
\begin{eqnarray}
\hat{\rho}(\mathbf{r},t) &\equiv& \hat{\rho}_s(\mathbf{r},t) +\hat{\rho}_l(\mathbf{r},t) \enskip ;\\
\hat{\mathbf{g}}(\mathbf{r},t) &\equiv& \hat{\mathbf{g}}_s(\mathbf{r},t)+\hat{\mathbf{g}}_l(\mathbf{r},t) \enskip ,
\end{eqnarray}
where $\hat{\rho}(\mathbf{r},t)=m \sum_i \delta[\mathbf{r}-\mathbf{r}_i(t)]$ and $\hat{\mathbf{g}}(\mathbf{r},t)=\sum_i \mathbf{p}_i(t) \delta[\mathbf{r}-\mathbf{r}_i(t)]$ are the total mass and momentum density of the system, respectively. The exact dynamical equations for the densities read (for the details of the derivation, see Appendix A):
\begin{eqnarray}
\label{MrhoS}\partial_t \hat{\rho}_s + \nabla\cdot\hat{\mathbf{g}}_s &=& +\hat{\sigma} \enskip ; \\
\label{MrhoL}\partial_t \hat{\rho}_l + \nabla\cdot\hat{\mathbf{g}}_l &=& -\hat{\sigma} \enskip ; \\
\label{MmomS}\partial_t \hat{\mathbf{g}}_s + \nabla\cdot\hat{\mathbb{K}}_s &=& -\hat{\rho}_s \nabla (\delta_{\hat{\rho}} \hat{V}) + \hat{\mathbf{j}} \enskip ; \\
\label{MmomL}\partial_t \hat{\mathbf{g}}_l + \nabla\cdot\hat{\mathbb{K}}_l &=& -\hat{\rho}_l \nabla (\delta_{\hat{\rho}} \hat{V}) - \hat{\mathbf{j}} \enskip ,
\end{eqnarray}
where $\hat{\sigma}(\mathbf{r},t)=m \sum_i \dot{\phi}_i(t) \delta[\mathbf{r}-\mathbf{r}_i(t)]$ is the microscopic local phase-change rate, $\hat{\mathbf{j}}(\mathbf{r},t)=\sum_i \dot{\phi}_i(t) \mathbf{p}_i(t)\delta[\mathbf{r}-\mathbf{r}_i(t)]$ is a microscopic momentum current density, $\delta_{\hat{\rho}} \hat{V} = \delta \hat{V}/\delta \hat{\rho}=\frac{v}{m^2}* \hat{\rho}$ is the first functional derivative of the potential energy of the system with respect to the microscopic mass density (the asterisk stands for spatial convolution), while $\hat{\mathbb{K}}_s(\mathbf{r},t)=\frac{1}{m}\sum_i \phi_i(t) [ \hat{\mathbf{p}}_i(t) \otimes \hat{\mathbf{p}}_i(t)] \delta[\mathbf{r}-\mathbf{r}_i(t)]$ and $\hat{\mathbb{K}}_l(\mathbf{r},t)=\frac{1}{m}\sum_i [1-\phi_i(t)] [ \hat{\mathbf{p}}_i(t) \otimes \hat{\mathbf{p}}_i(t)] \delta[\mathbf{r}-\mathbf{r}_i(t)]$ {(where $\otimes$ stands for the dyadic product)} are the microscopic kinetic stress tensors of the solid and liquid, respectively, which also indicates $\hat{\mathbb{K}}(\mathbf{r},t) \equiv \hat{\mathbb{K}}_s(\mathbf{r},t)+\hat{\mathbb{K}}_l(\mathbf{r},t)$, where $\hat{\mathbb{K}}(\mathbf{r},t)=\frac{1}{m}\sum_i [ \hat{\mathbf{p}}_i(t) \otimes \hat{\mathbf{p}}_i(t)] \delta[\mathbf{r}-\mathbf{r}_i(t)]$ is the kinetic stress tensor. To account for phase transition, we keep Eq. (\ref{MrhoS}), but add Equations (\ref{MrhoS}) and (\ref{MrhoL}), and (\ref{MmomS}) and (\ref{MmomL}), which results in:
\begin{eqnarray}
\label{Mphi}\partial_t \hat{\rho}_s + \nabla\cdot\hat{\mathbf{g}}_s &=& \hat{\sigma} \enskip ; \\
\label{Mcont}\partial_t \hat{\rho} + \nabla\cdot\hat{\mathbf{g}} &=& 0 \enskip ; \\
\label{MNS}\partial_t \hat{\mathbf{g}} + \nabla\cdot\hat{\mathbb{K}} &=& -\hat{\rho} \nabla (\delta_{\hat{\rho}} \hat{V}) \enskip ,
\end{eqnarray}
where the first equation accounts for the phase transition, the second for mass continuity, while the third is a microscopic Navier-Stokes equation. To proceed, we define our macroscopic variables as the ensemble average of their microscopic counterparts over the initial condition \cite{PhysRevLett.112.100602}: 
\begin{equation}
\label{coarse}Y(\mathbf{r},t):=\int d\mathbb{P}_0(\mathbf{\Gamma}) \hat{Y}(\mathbf{r},t) \enskip ,
\end{equation} 
where $Y=\rho,\mathbf{g},\mathbb{K}$ and $\sigma$. {We note that the above expression coincides with the more commonly used $Y(\mathbf{r},t) := \int d\mathbb{P}(\mathbf{\Gamma},t)\hat{Y}(\mathbf{r})$ [where $\hat{Y}(\mathbf{r})$ is the phase-space operator of the quantity] for $d\mathbb{P}(\mathbf{\Gamma},0)=d\mathbb{P}_0(\mathbf{\Gamma})$. This equivalence can be proven by using a coordinate transformation \cite{KhinchinGamow} and the fact that the phase-space probability density $f(\mathbf{\Gamma},t)$ is driven by the Liouville equation.} Applying Eq. (\ref{coarse}) to Equations (\ref{Mphi})-(\ref{Mcont}) results in:
\begin{eqnarray}
\label{MRphi}\partial_t \rho_s + \nabla\cdot(\rho_s \mathbf{v}) &=& \sigma \enskip ; \\
\label{MRcont}\partial_t \rho + \nabla\cdot(\rho\,\mathbf{v}) &=& 0 \enskip , 
\end{eqnarray}
where we introduced the macroscopic velocity field $\mathbf{v}(\mathbf{r},t):=\mathbf{g}(\mathbf{r},t)/\rho(\mathbf{r},t)$, and {set
\begin{equation*}
\mathbf{g}/\rho=\mathbf{g}_s/\rho_s = \mathbf{g}_l/\rho_l \enskip ,
\end{equation*}
which postulates that the solid and liquid phases are \textit{co-moving} on the macroscopic level. The explanation of this principle starts on the microscopic level:} The microscopic mass densities are invariant for the spatial permutation of the particles, simply because the bond-order parameter merely depends on the structure of the neighbourhood of a particle. Consequently, swapping two {identical,} neighbouring particles (this is the elementary process of diffusion between two particles of different types in a binary system) results in exactly the same microscopic densities, {and therefore no diffusion is expected} between the solid and liquid phases (see Fig. 1).
\begin{figure}
\includegraphics[width=1.0\linewidth]{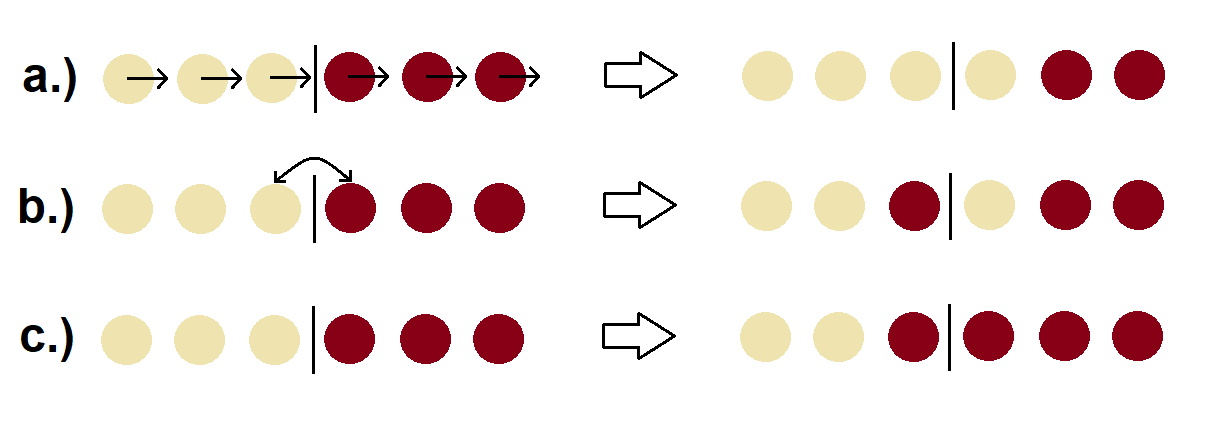}
\caption{Schematic illustration of elementary processes in a classical many-particle system: a.) Convection (collective movement of particles), b.) Diffusion (relative movement of particles of different species), and c.) Phase transition (change in the structure of the local environment). Note that a.) and b.) are elementary transport modes, while c.) is not associated with the movement of the particle that changed colour.}
\end{figure}
{Alternatively, since every single particle contributes to both phases, the phases consist of the same particles. Since a particle cannot move relative to itself, we don't expect diffusion on the macroscopic level. Summarising,} the only transport mode in a single component material is convection, while the phase transition between the solid and a liquid is a local structural change driven by $\sigma$. To complete the set of dynamical equations, finally we need to apply Eq. (\ref{coarse}) to Eq. (\ref{MNS}) as well, which results in the following well-known Navier-Stokes equation:
\begin{equation}
\label{Euler}\partial_t (\rho \mathbf{v}) + \nabla\cdot(\rho\,\mathbf{v} \otimes \mathbf{v}) = -\rho \nabla (\delta_{\rho} F) \enskip , 
\end{equation}
where $F[\rho]=\int dV\,f(\rho,\nabla\rho,\nabla^2\rho,\dots)$ is the Helmholtz free energy of the non-equilibrium system (a functional of $\rho$), and we used two standard approximations \cite{doi:10.1063/1.4913636}, (i) the ideal gas approximation $\mathbb{K} \approx \rho \, \mathbf{v} \otimes \mathbf{v} + \left(\frac{k_B T}{m}\right) \rho$, and (ii) the adiabatic approximation $\int d\mathbb{P}(\mathbf{\Gamma}_0) \{\hat{\rho} \nabla (\delta_{\hat{\rho}} \hat{V})\} \approx \rho\nabla(\delta_\rho V)$, where $V = \frac{1}{2\, m^2}\int dV \{\rho(v * \rho)\}$ is the coarse-grained potential energy. Before proceeding to constructing $\sigma$ and the free energy functional, we need to address a major issue respecting Eq. (\ref{Euler}). Since no external force is imposed on the system, the full momentum of the system must be conserved, which applies to Eq. (\ref{MNS}) (see Appendix A). Since the coarse-graining operator described by Eq. (\ref{coarse}) doesn't operate on $t$ and $\mathbf{r}$, the total coarse-grained momentum is also conserved. This indicates, that $\int d\mathbb{P}(\mathbf{\Gamma}_0) \{\hat{\rho} \nabla (\delta_{\hat{\rho}} \hat{V})\}$ and $\mathbb{K}$ must be approximated in non-equilibrium only in such ways, which guarantee the conservation of the coarse-grained momentum. A thermodynamically consistent condition for this is, that there must exist such a tensor $\mathbb{R}$ (called reversible stress) being the explicit function of the coarse-grained variables and their spatial derivatives, whose divergence equals the right-hand side of Eq. (\ref{Euler}). As long as the coarse-grained density fully characterises the system, the right-hand side of Eq. (\ref{eq3}) is valid, since such a tensor exists \cite{doi:10.1063/1.432687,PhysRevLett.109.120603,Goddard_2012,doi:10.1063/1.4913636,doi:10.1063/1.4807586,doi:10.1063/1.5008608}. In contrast, we have two coarse-grained variables characterising the spatio-temporal state of the system, and therefore there's no guarantee that such an $\mathbb{R}$ exists, for which $\nabla\cdot\mathbb{R}=-\rho\nabla(\delta_\rho F[\rho_s,\rho_l])$ for arbitrary $F[\rho_s,\rho_l]$. To "fix" Eq. (\ref{Euler}), we re-define the right-hand side of Eq. (\ref{Euler}) as (see Appendix B):
\begin{equation}
\label{GD}
\partial_t (\rho\,\mathbf{v}) + \nabla\cdot(\rho\,\mathbf{v} \otimes \mathbf{v}) = -\rho_s\nabla(\delta_{\rho_s}F)-\rho_l\nabla(\delta_{\rho_l}F) \enskip .
\end{equation}  
The above equation preserves the full momentum of the system if such a tensor exists, whose divergence coincides with the right-hand side of Eq. (\ref{GD}). Assuming that the free energy functional depends only on the solid and liquid densities and their gradients (Landau theory of first order phase transitions), the tensor reads \cite{Korteweg1901}: $\mathbb{R} = -p\,\mathbb{I} + \mathbb{A}$, where $-p = f -\rho_s (\delta_{\rho_s} F) + \rho_l (\delta_{\rho_l} F)$ and $\mathbb{A} = - \nabla\rho_s \otimes (\partial_{\nabla\rho_s}f) - \nabla\rho_l \otimes (\partial_{\nabla\rho_l}f)$.

\subsubsection{Phase-change rate and variable transformation}

The next step of the derivation of the general dynamical equations is postulating the phase-change rate in Eq. (\ref{MRphi}). Assuming that $F$ can be expressed in terms of $\rho_s$ and $\rho_l$, and taken into account that $\frac{1}{V}\int dV (\rho_s+\rho_l)=\bar{\rho}$ (constant) result in the following Euler-Lagrange equations:
\begin{equation}
\label{ELrho}
\delta_{\rho_s} F  = \delta_{\rho_l} F = \mu \enskip ,
\end{equation}
where $\mu$ is constant. Note that the right-hand side of Eq. (\ref{GD}) becomes $0$ for Eq. (\ref{ELrho}), while the phase change rate can be postulated as \cite{PhysRevB.8.3423}:
\begin{equation}
\label{rate}
\sigma := -M (\delta_{\rho_s}F - \delta_{\rho_l}F) \enskip ,
\end{equation}
which also becomes $0$ in equilibrium. Using the variables $c_s:=\rho_s/(\rho_s+\rho_l)$, $c_l:=\rho_l/(\rho_s+\rho_l)$ and $\rho=\rho_s+\rho_l$ (local mass fractions of the phases and the total local mass density, respectively) in Equations (\ref{MRphi}), (\ref{MRcont}) and (\ref{GD}) results in the following dynamical equations (for the mathematical details of the variable transformation, see Appendix B):
\begin{eqnarray}
\label{eq1} \rho \, \dot{\phi} &=& -(M/\rho)(\delta_\phi F) \\
\label{eq2} \dot{\rho} &=& -\rho(\nabla\cdot\mathbf{v}) \\
\label{eq3} \rho \, \dot{\mathbf{v}} &=& -\rho\nabla(\delta_\rho F)+(\delta_\phi F)\nabla\phi \enskip ,
\end{eqnarray}  
where the phase field $\phi(\mathbf{r},t) \equiv c_s(\mathbf{r},t)$ is the local solid mass fraction, $\dot{y}=\partial_t y + \mathbf{v}\cdot \nabla y$ stands for the material derivative. Furthermore, Eq. (\ref{ELrho}) transforms as:
\begin{equation}
\label{ELtrans} \delta_\phi F = 0 \quad \text{and} \quad \delta_\rho F = \mu \enskip ,
\end{equation}
and therefore any equilibrium solution represents stationary states of the dynamical equations (\ref{eq1})-(\ref{eq3}), which now completely define the time evolution of a single component, two-phase system for a given $F[\phi,\rho]$. Applying the variable transformation to the reversible stress tensor results in $-p = f - \rho(\delta_\rho F)$ and $\mathbb{A} = -\nabla\rho\otimes(\partial_{\nabla\rho f})-\nabla\phi\otimes(\partial_{\nabla\phi f})$, which indeed recovers
\begin{equation}
\label{trueGD}
\nabla\cdot\mathbb{R} = -\rho\nabla(\delta_\rho F)+(\delta_\phi F)\nabla\phi \enskip .
\end{equation}
Finally we mention, that Eq. (\ref{eq3}) accords with the results of Anderson, McFadden and Wheeler \cite{Wheeler08081997,doi:10.1146/annurev.fluid.30.1.139,Anderson2000175}, and Conti \cite{PhysRevE.64.051601,PhysRevE.69.022601}, but differs from Eq. (\ref{Euler}) used by Oxtoby and Harrowell \cite{doi:10.1063/1.462864}. While it is straightforward to find the stress tensor for Eq. (\ref{eq3}), it exist for Eq. (\ref{Euler}) for only those $F[\rho,\phi]$, where the phase-field and the density are decoupled (see Appendix B), {which applies to the model of Oxtoby and Harrowell, and therefore their approach is also acceptable.}

\subsubsection{Free energy functional}

The next step of the model development is to construct a free energy functional. For the sake of mathematical simplicity we choose
\begin{equation}
\label{func} F[\phi,\rho] := F_0[\phi] + K \int dV f_\rho(\phi,\rho/\rho_0) \enskip ,
\end{equation}
where $F_0[\phi]$ is the standard phase-field model of solidification \cite{Langer,Provatas:2010:PMM:1965378}:
\begin{equation}
\label{func0}
F_0[\phi] := \int dV \left\{ \frac{3\,\sigma\,\delta}{2}|\nabla\phi|^2 + \frac{\sigma}{\delta}\,g(\phi) + \Lambda \, p(\phi) \right\} \enskip ,
\end{equation} 
where 
\begin{eqnarray*}
g(\phi)&=&6 \, [\phi(1-\phi)]^2 \enskip ; \\
p(\phi)&=&\phi^2(3-2\,\phi) \enskip ,
\end{eqnarray*}
and $\Lambda$ is the thermodynamic driving force for solidification. Furthermore, $f_\rho(\phi,\varrho)$ is a local (dimensionless) coupling between the phase-field and the density (to be defined later), and $K$ the bulk modulus of the system. For $f_\rho(\phi,\varrho)\equiv 0$, $\sigma$ and $\delta$ [in Eq. (\ref{func0})] are the free energy and characteristic width of the equilibrium ($\Lambda=0$) planar crystal-liquid interface, respectively. We also postulate
\begin{equation}
\label{PFmob}
M := \Gamma(\phi) \, \rho \enskip ,
\end{equation} 
where $\Gamma(0)=\Gamma(1)=\Gamma_0$ (constant), which is necessary for the existence of the propagating steady-state front solution (we will show this later). Finally, we non-dimensionalise the system by setting the length scale to $\delta$, the density scale to $\rho_0$, and the energy scale to $\sigma\delta^3$, which results in:
\begin{equation}
\label{scaledF}
F[\phi,\rho] = F_0[\phi]+ B \int dV f_\rho(\phi,\rho) \enskip ,
\end{equation}
where
\begin{equation*}
F_0[\phi] = \int dV \left\{ \frac{3}{2}(\nabla\phi)^2 + g(\phi)+ \lambda \, p(\phi) \right\} \enskip ,
\end{equation*}
while the global condition for the density reads:
\begin{equation*}
\bar{\rho} = \frac{1}{V} \int dV \rho \equiv 1 \enskip .
\end{equation*}
The dimensionless local free energy density for different density-phase couplings is shown in Fig 2. Furthermore, choosing the time scale $\sqrt{\bar{\rho}\,\delta^3/\sigma}$ in Equations (\ref{eq1})-(\ref{eq3}) yields:
\begin{eqnarray}
\label{neq1} \rho \,\dot{\phi} &=& -\kappa(\delta_\phi F) \\
\label{neq2} \dot{\rho} &=& -\rho(\nabla\cdot\mathbf{v}) \\
\label{neq3} \rho \, \dot{\mathbf{v}} &=& -\rho \nabla(\delta_\rho F) + (\delta_\phi F)\nabla \phi \enskip ,
\end{eqnarray}  
where the functional derivatives read:  
\begin{eqnarray}
\label{varderphi}\delta_\phi F &=& \delta_\phi F_0[\phi] + B \, \partial_\phi f_\rho(\phi,\rho) \enskip ; \\
\label{varderrho}\delta_\rho F &=& B \, \partial_\rho f_\rho(\phi,\rho) \enskip ,
\end{eqnarray}
where $\delta_\phi F_0[\phi] = g'(\phi)+\lambda\,p'(\phi)-3\nabla^2\phi$ {(where $'$ stands for the derivative with respect to the argument)}. The dimensionless model parameters read: $\lambda=(\delta/\sigma)\Lambda$, $B=(\delta/\sigma)K$, and $\kappa=\Gamma \sqrt{\sigma\delta/\bar{\rho}}$.

\subsubsection{Quasi-incompressible hydrodynamics}

Equations (\ref{neq1})-(\ref{neq3}) contain two time scales associated with (i) solidification (the corresponding model parameters are $\kappa$ and $\lambda$), and (ii) sound waves (via $b$). Let now $\lambda:=0$ (equilibrium) and let $\kappa$ be constant. In the \textit{gapless} limit, the density dependent part of the free energy density can be defined as:
\begin{equation}
\label{gapless}
f_\rho(\phi,\rho):=(1/2)(\rho-1)^2 \enskip ,
\end{equation} 
which indicates that the density and the phase-field are decoupled. Solving the Euler-Lagrange equations for $\lambda=0$ gives the following bulk equilibrium phases: $\phi=0$, $\phi=1$, and $\phi=1/2$, with $\rho=\bar{\rho}$. The linearisation of Equations (\ref{neq1})-(\ref{neq3}) around $\phi=0$ (or $\phi=1$), $\rho=1$ and $v=0$ (bulk equilibrium phases) in one spatial dimension yields:
\begin{equation}
\label{linear_comp}
\partial_t \left( \begin{matrix} \delta \phi \\ \delta\rho \\ \delta v \end{matrix} \right) = \left( \begin{matrix}  -3\kappa(4-\partial_x^2) & 0 & 0 \\ 0 & 0 & -\partial_x  \\ 0 & -B \, \partial_x & 0 \end{matrix} \right) \left( \begin{matrix} \delta \phi \\ \delta\rho \\ \delta v \end{matrix} \right) \enskip .
\end{equation}
This system is decoupled in the phase-field and the density-velocity pair, and the dispersion relations read:
\begin{eqnarray}
\label{dispersion1}\omega_\phi(k)&=&-3\,\kappa\,(4+k^2) \\
\label{dispersion2}\omega_{(\rho,v)}(k) &=& \pm\imath\,k\,\sqrt{B} \enskip .
\end{eqnarray}
Eq. (\ref{dispersion1}) indicates that the phase field is stable around $\phi=0$ and $\phi=1$, and the long wavelength perturbations relax on the characteristic time scale $1/(12\kappa)$, while the density-velocity dispersion relation indicates the presence of sound waves, where the speed of sound is $c=\sqrt{B}$. [Here we mention, that the linearisation around $\phi=1/2$ would give $\text{Re}[\omega_\phi(0)]>0$, thus indicating an unstable intermediate phase.] The characteristic times scale of pattern formation can be estimated by using the Wilson-Frenkel model \cite{doi:10.1080/14786440009463908,Frenkel} in the small driving force limit, yielding (see Appendix C):
\begin{equation*}
\kappa \approx\frac{D a_0}{3\,\ell^2} \frac{v_M}{R\,T}\sqrt{\frac{\sigma\,\bar{\rho}}{\delta}} \enskip ,
\end{equation*}
where $D$ is the self-diffusion coefficient, $a_0$ the inter-atomic spacing, $\ell$ the diffusional mean free path in the liquid, $v_M$ the molar volume, $T$ the temperature, and $R$ the universal gas constant. Taking typical values for liquid metals, i.e., $\sigma \approx 1$ J/m$^2$, $\delta \approx 1$ nm, $K \approx 100$ GPa (crystal), $\bar{\rho}=5000$ kg/m$^3$, $D=3.5 \cdot 10^{-9}$ m$^2$/s, $v_M=10^{-5}$ m$^3$/mol, $T=1500K$, $a_0 \approx (v_M/N_A)^{1/3}$, $\ell = 2D/\bar{v}$ (where $\bar{v}=\sqrt{3 k_B T/m}$ is the average velocity of the particles, and $m=\bar{\rho}\,v_M/N_A$ is the particle mass) results in 
\begin{equation}
\kappa \approx 10 \quad \text{and} \quad c \approx 10 \enskip .
\end{equation}
Since the dimensionless front speed reads $V = -3\,\kappa\,\lambda$ (see Appendix C), and the theory applies to $|\lambda| \ll 1$, the solidification is typically much slower than sound waves, and is therefore inaccessible in numerical simuations. This problem emerged in the works of Conti \cite{PhysRevE.64.051601,PhysRevE.69.022601}, and was overcome by reducing the bulk modulus. Since the speed of sound is proportional to the square root of the modulus, an order of magnitude reduction in the speed of sound necessitates two orders of magnitude reduction in the modulus, while two orders of magnitude reduction in $c$ would necessitate 4 orders of magnitude reduction in $B$, and so on. Unfortunately, the reduction of $B$ significantly changes the equation of state, and also might affect the interface behaviour. To resolve the problem, here we are proposing a different approach: We will eliminate sound waves, together with allowing the density to vary across the crystal-liquid interface. The physical interpretation of this step is that we consider the system incompressible in the sense that sound waves relax instantaneously, which corresponds to $B \to \infty$, which is physically more justifiable than modelling the solid-liquid system as a (highly) compressible gas. The stationary solutions of the dynamical equations can be determined by solving the following Euler-Lagrange equations [emerging from Equations (\ref{ELtrans}), (\ref{varderphi}) and (\ref{varderrho})]:
\begin{eqnarray}
\label{planar1} \delta_\phi F_0[\phi] + B \, \partial_\phi f_\rho(\phi,\rho) &=& 0 \\
\label{planar2} B \, \partial_\rho f_\rho(\phi,\rho) &=& \mu \enskip ,
\end{eqnarray}
where the second equation is only an algebraic equation locally connecting $\phi$ and $\rho$. Now we require the followings:
\begin{itemize}
\item A unique planar interface solution exists in equilibrium, which connects the stable homogeneous phases $\phi=1$ in $x \to -\infty$ and $\phi=0$ in $x \to +\infty$.
\item The density is an explicit function of the phase-field at the equilibrium planar interface solution: $\rho_0(x)=h[\phi_0(x)]$.
\end{itemize}
To eliminate sound waves, we {generalise} the idea of Truskinowsky and Lowengrub \cite{Lowengrub2617}, and postulate the local condition
\begin{equation}
\label{lcond} \rho(\mathbf{r},t) := h[\phi(\mathbf{r},t)] \enskip ,
\end{equation} 
which means that the density is {locally set by the phase. (Our general understanding of \textit{quasi-incompressibility} is that the total local density of the system is an explicit function of the other state variables.)} This condition can be taken into account in Equations (\ref{neq1})-(\ref{neq3}) by using the Lagrange multiplier method, thus yielding the following conditional hydrodynamic equations (see Appendix D):
\begin{eqnarray}
\label{qeq1} h \,\dot{\phi} &=& -\kappa \left( \delta_\phi^{(c)}F \right) \\
\label{qeq2} \nabla\cdot\mathbf{v} &=& -\frac{1}{h}\frac{dh}{d\phi}\dot{\phi} \\
\label{qeq3} h \,\dot{\mathbf{v}} &=& \left( \delta_\phi^{(c)}F \right) \nabla \phi - \nabla \delta p \enskip ,
\end{eqnarray}
where $\delta_\phi^{(c)} F = (\delta_\phi F)_{\rho=h(\phi)}$ {[the functional derivative evaluated at $\rho=h(\phi)$]} is the "conditional functional derivative" of $F$ with respect to $\phi$, and $\delta p(\mathbf{r},t)$ is responsible for Eq. (\ref{qeq2}). Note that the equilibrium planar interface solution is a stationary solution of Equations (\ref{qeq1})-(\ref{qeq3}) with $\delta p=0$. Finally we mention, that Eq. (\ref{qeq3}) preserves the full momentum of the system, since its right-hand side is identical to $\nabla \cdot \left( \mathbb{R}|_{\rho=h(\phi)}- \delta p\,\mathbb{I} \right)$, where $\mathbb{R}=[f - \rho (\delta_\rho F)]\mathbb{I}-\nabla\phi \otimes (\partial_{\nabla\phi} f)$ is the compressible reversible stress. The most straightforward scenario to test the quasi-incompressible hydrodynamics against sound waves is the gapless limit [see Eq. (\ref{gapless})]. Here the system is incompressible, since the Euler-Lagrange equation for the density has the unique solution $\rho_0(x) = 1$, since $\bar{\rho} \equiv 1$. The linearisation of Equations (\ref{qeq1})-(\ref{qeq3}) around $\phi=0$ (or $\phi=1$), $v=0$ and $\delta p=0$ results in:
\begin{eqnarray}
\nonumber \partial_t \delta \phi(x,t) &=& -\kappa (12-3\partial_x^2)\delta\phi(x,t) \\
\label{cstab2}\partial_x \delta v(x,t) &=& 0  \\
\nonumber \partial_t \delta v(x,t) &=& - \partial_x \delta p(x,t) \enskip .
\end{eqnarray}
The above system is similar to Eq. (\ref{linear_comp}) in the sense that it is decoupled in the phase-field and the density-velocity, and prescribes the same behaviour for the relaxation of the phase-field. The major difference is, that Eq. (\ref{cstab2}) indicates $\delta v(x,t) = 0$ (incompressibility), which also results in $\delta p(x,t)=0$, and therefore the time scale associated with $B$ is eliminated. alternatively, one can say that the equation of state, $f_\rho(\phi,\rho)$ is "replaced" by the pressure correction $\delta p(\mathbf{r},t)$.

\subsection{Planar interfaces and solidification front speed}

\subsubsection{Equilibrium planar solid-liquid interface}

The density-phase relationship appearing in the quasi-incompressible hydrodynamic equations can be found by solving the Euler-Lagrange equations for the equilibrium ($\lambda=0$) planar interface satisfying the boundary conditions $\lim_{x \to -\infty}\phi(x)=1$ and $\lim_{x \to +\infty}\phi(x)=0$. The interface connects the two homogeneous phases, whose densities can be determined by solving the following equations (often called the common tangent construction, see Appendix E): 
\begin{eqnarray}
\label{commtan1} \mu_s(\rho_s^0) &=& \mu_l(\rho_l^0) \enskip ;  \\
\label{commtan2} f_s(\rho_s^0) - \rho_s^0 \mu_s(\rho_s^0) &=& f_l(\rho_l^0) - \rho_l^0 \mu_l(\rho_l^0) \enskip ,
\end{eqnarray}   
where $f_s(\rho) = f_\rho(1,\rho)$ and $f_l(\rho) = f_\rho(0,\rho)$ are the free energy densities of the bulk solid and liquid phases, respectively, while $\mu_{s,l}(\rho)=df_{s,l}(\rho)/d\rho$ are the chemical potentials. Since we have 2 equations for 2 unknowns, the equilibrium densities are fixed, and therefore the equilibrium chemical potential $\mu_0:= \mu_s(\rho_s^0)= \mu_l(\rho_l^0)$ is also fixed by the construction. Introducing $\epsilon := (\rho_s^0-\rho_l^0)/(\rho_s^0+\rho_l^0)$, the half of the relative coexistence density gap, results in $\rho_l^0 = \rho_c(1-\epsilon)$ and $\rho_s^0=\rho_c(1+\epsilon)$, where $\rho_c=(\rho_s^0+\rho_l^0)/2$. In our work, we always scale the dimensional density by $\rho_0:=\rho_c$, which result in $\bar{\rho} \equiv 1$, and therefore the equilibrium planar interface connects the bulk phases $(\phi,\rho)=(0,1-\epsilon)$ and $(1,1+\epsilon)$. Finally we mention, that the stability of the homogeneous bulk phases must always be checked in the quasi-incompressible dynamics to ensure, that the only stable interface is the one connecting the bulk liquid at density $1-\epsilon$ to the bulk solid at density $1+\epsilon$.
 
\subsubsection{Propagating 1-dimensional solidification front}

\paragraph{Compressible hydrodynamics.} The next step is to check whether Equations (\ref{neq1})-(\ref{neq3}) provide a one-dimensional propagating steady interface solution. In a propagating steady solution any field reads: $\varphi(x,t) = \varphi_0(z)$, where $z=x-V t$, where $V$ is the front velocity in the frame of reference where the solid stands still at $x \to -\infty$, thus indicating the boundary condition $\lim_{x \to -\infty}v(x,t):=0$. Using the transformation in Equations (\ref{neq1})-(\ref{neq3}) yields:
\begin{eqnarray}
\label{intf1} \rho(v-V)\phi' &=& -\kappa (\delta_\phi F) \\
\label{intf2} (v-V)\rho' &=& -\rho \, v' \\
\label{intf3} \rho(v-V)v' &=& \phi' (\delta_\phi F) - \rho (\delta_\rho F)' \enskip ,
\end{eqnarray}  
where $(.)'$ indicates differentiation with respect to $z$. The particular solution of Eq. (\ref{intf2}) for $\lim_{z \to -\infty} v(z)=0$ and $\lim_{z \to -\infty} \rho(z):=1+\Delta$ reads:
\begin{equation}
\label{vprofile}
v(z) = V \left[ 1 - \frac{1+\Delta}{\rho(z)}\right] \enskip .
\end{equation}
Using Eq. (\ref{vprofile}) in Equations (\ref{intf1}) and (\ref{intf3}), then integrating the latter with respect to $z$ from $-\infty$ yields:
\begin{eqnarray}
\label{comp1} W\,\phi' &=& \kappa (\delta_\phi F) \\
\label{comp2} W\,v &=& \lambda-R \enskip ,
\end{eqnarray} 
where $W = (1+\Delta)V$ and $R = f - \rho(\partial_\rho f)-\phi'(\partial_{\phi'}f)$. The above eqautions are then to be solved for $\phi(z)$ and $\rho(z)$ for a particular $f_\rho(\phi,\rho)$.\\

\paragraph{Quasi-incompressible hydrodynamics.} Applying the steps described in the previous section for Equations (\ref{qeq1})-(\ref{qeq3}) in the general case [$\epsilon \in (-1,1)$], the following equations emerge for the propagating planar interface:
\begin{eqnarray}
\label{cntf1} h(v-V)\phi' &=& -\kappa \left( \delta_\phi^{(c)} F \right) \\
\label{cntf2} v' &=& -\frac{1}{h}\frac{dh}{d\phi}(v-V)\phi' \\
\label{cntf3} h(v-V)v' &=& \phi' \left( \delta_\phi^{(c)} F \right) - \delta p' \enskip ,
\end{eqnarray}
Analogously to Eq. (\ref{vprofile}), the solution of Eq. (\ref{cntf2}) for $\lim_{z \to-\infty}h[\phi(z)]:=1+\Delta$ reads:
\begin{equation}
\label{cvprofile}
v(z) = V \left\{ 1 - \frac{1+\Delta}{h[\phi(z)]} \right\} \enskip .
\end{equation} 
Using Eq. (\ref{cvprofile}) in Equations (\ref{cntf1}) and (\ref{cntf3}) results in:
\begin{eqnarray}
\label{quasiPF} - W\,\phi' &=& -\kappa \left( \delta_\phi^{(c)} F \right) \\
 - W\,v' &=& \phi' \left( \delta_\phi^{(c)} F \right) - \delta p' \enskip ,
\end{eqnarray}  
where $W=(1+\Delta) V$. To check whether the equilibrium planar interface solution $\phi_0(x)$ represents a propagating interface solution, we replace $\Delta$ by $\epsilon$ and use $\phi(z)$ $\phi(z):=\phi_0(z)$ in the above equations. Recalling that $\kappa(\phi)|_{\phi=0,1}=\kappa_0$ (constant), and assuming that there exists $\kappa(\phi):=\kappa_0\,[1+\delta\kappa(\phi)]$ so that $\kappa[\phi_0(z)]p'[\phi_0(z)]/\phi_0'(z)$ is constant yield:
\begin{equation}
\label{frontV} V = \left( \frac{\lambda\,\kappa_0}{1+\epsilon} \right) \lim_{z \to \pm \infty}\left\{ \frac{p'[\phi_0(z)]}{\phi_0'(z)}\right\}  \enskip , \\
\end{equation}
where we used that $\delta_\phi^{(c)} F  = \lambda \, p'[\phi_0(z)]$ for the planar equilibrium interface solution, and $\lim_{z \to \pm \infty} \delta\kappa[\phi_0(z)] = \delta\kappa(\phi)|_{\phi=0,1}=0$. Finally, the pressure correction reads:
\begin{equation*}
\delta p(z) = \lambda \, p[\phi_0(z)] + (1+\epsilon)V\,v(z) \enskip .
\end{equation*}
We note here that the front velocity can be calculated \textit{without} determining the equilibrium interface, emerging from the fact that $\phi_0'(z)=\sqrt{(2/3)S[\phi_0(z)]}$ in Eq. (\ref{frontV}) (see Appendix E), thus yielding:
\begin{equation}
\label{frontVsimple}
\lim_{z \to \pm \infty}\left[ \frac{p'[\phi_0(z)]}{\phi'_0(z)} \right] = \sqrt{\frac{3}{2}}\lim_{\phi \to 0,1}\left[ \frac{p'(\phi)}{\sqrt{S(\phi)}}\right] \enskip ,
\end{equation}
where $S(\phi) = g(\phi) + B \int_{0}^\phi d\psi \left\{ [\partial_\psi f_\rho(\psi,\rho)]_{\rho=h(\psi)} \right\}$. Eq. (\ref{frontVsimple}) immediately indicates
\begin{equation}
\label{kappaMOD}
\delta\kappa(\phi) = \chi \frac{\sqrt{S(\phi)}}{p'(\phi)} -1 \enskip ,
\end{equation}
where $\chi = \lim_{\phi \to 0,1}[p'(\phi)/\sqrt{S(q)}]$. Finally we mention, that Eq. (\ref{PFmob}) is a necessary condition for the existence of $V$. Since $\lim_{z \to \pm \infty}\left[ p'[\phi_0(z)]/\phi'_0(z) \right]$ (or, equivalently, $\lim_{\phi \to 0,1} [ p'(\phi)/\sqrt{S(\phi)}]$) is \textit{unique}, $V$ exists if and only if the pre-factor of the limes is constant in Eq. (\ref{frontV}). If $\Gamma$ was not proportional to $\rho$, the pre-factor would contain $1/\rho$, which differs in the limits $z \to \pm \infty$.

\section{Interface analysis for different density-phase couplings}

\begin{figure}
\includegraphics[width=0.49\linewidth]{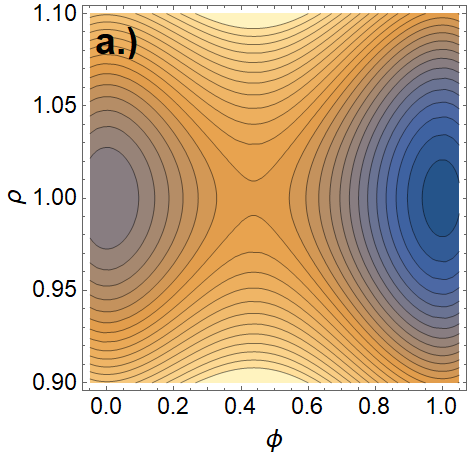}
\includegraphics[width=0.49\linewidth]{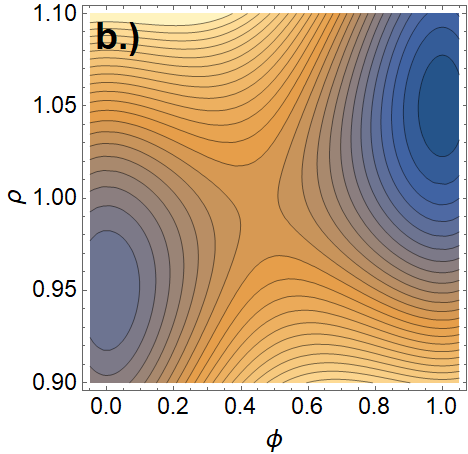}
\includegraphics[width=0.49\linewidth]{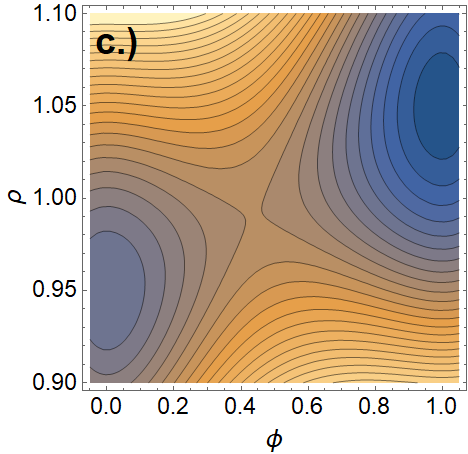}
\includegraphics[width=0.49\linewidth]{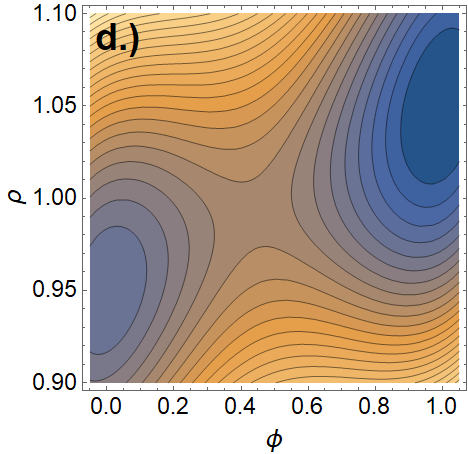}
\caption{Contour plot of the local free energy density in a.) the gapless model [Eq. (\ref{Model0})], b.) Model 1 [Eq. (\ref{Model1})], c.) Model 2 [Eq. (\ref{Model2})] for $q(\phi)=1+\epsilon[2\,p(\phi)-1]$, and d.) Model 2 [Eq. (\ref{Model2})] for $q(\phi)=1/[\phi/(1+\epsilon)+(1-\phi)/(1-\epsilon)]$. The parameters are $B=100$, $\epsilon=0$ for panel a.) and $\epsilon=0.05$ for panels b.)-d.), and $\lambda=-0.25$. Darker shades indicate lower values. Note the two minima at $\phi,\rho=(0,1-\epsilon)$ and $(1,1+\epsilon)$, and the saddle point between them.}
\end{figure} 

\subsection{Model 0}

In the gapless limit, the density dependent part of the free energy density read:
\begin{equation}
\label{Model0}
f_\rho(\phi,\rho):=\frac{(\rho-1)^2}{2} \enskip ,
\end{equation}
which indicates that the phase-field and the density are decoupled [see Fig 2(a)]. The Euler-Lagrange equations read:
\begin{eqnarray}
\label{nogapphi} (dg/d\phi)|_{\phi=\phi_0(x)} - 3\phi_0''(x) &=& 0 \\
\label{nogaprho} B [\rho_0(x)-1] &=& \mu_0 \enskip .
\end{eqnarray} 
The equations have 3 homogeneous solutions at arbitrary density, namely $\phi=0,1$ and $1/2$. The common tangent construction prescribes $\rho_s=\rho_l=1$ and $\mu_0=0$ for $\bar{\rho}=1$, thus indicating no density gap between the coexisting bulk equilibrium liquid and solid phases. Finally, Eq. (\ref{nogapphi}) provides the well-known interface solution
\begin{equation}
\label{basicprofile}
\phi_0(x) = \frac{1-\tanh(x)}{2} \enskip .
\end{equation}
It is worth to investigate whether Eq. (\ref{basicprofile}) represents a propagating planar interface of the compressible dynamics. Since $\delta_\phi F=\delta_\phi F_0$ (no density contribution), the exact solution of Eq. (\ref{comp1}) for $\kappa=\kappa_0$ (constant) reads: $\phi(z)=[1-\tanh(z)]/2 \equiv \phi_0(z)$ with
\begin{equation}
V = - \frac{3\, \kappa_0 \, \lambda}{1+\Delta} \enskip .
\end{equation}
Using these in Eq. (\ref{comp2}), and taking the $z \to +\infty$ limit results in:
\begin{equation}
(6\,\kappa_0\,\lambda) ^2\Delta = (B\,\Delta+\lambda) (4-\Delta^2) \enskip .
\end{equation}
The solution of the above equation reads:
\begin{equation}
\label{compgapless}
\Delta = \frac{\lambda}{2\,B} + O(\lambda^3)\enskip ,
\end{equation}
which indicates that neither the density is constant nor the velocity field is zero for a propagating  planar interface in the gapless model. More precisely, a dynamical density gap $\rho_s-\rho_l = 2\,\Delta \approx \lambda/B<0$ appears for $\lambda<0$ and $|\lambda|\ll 1$, which indicates \textit{expansion} upon solidification, however, it is negligible, as long as $\lambda \ll B$. It is worth to mention, that the dynamical density gap emerges from the contribution of the phase-field to the reversible stress tensor. Since $\rho$ and $\phi$ are decoupled on the level of the free energy functional in the gapless limit, the right-hand side of the Navier-Stokes equation could be re-defined as $ -\rho \nabla(\delta_\rho F)$. The corresponding reversible stress reads $\mathbb{R} = \left[ f_\rho - \rho (d_\rho f_\rho) \right] \, \mathbb{I}$, for which $\nabla\cdot\mathbb{R}=-\rho \nabla(\delta_\rho F)$, and therefore the full momentum of the system is constant. This modification would result in $\rho(z)=1$ and $v(z)=0$ for the propagating interface, but one should remember, that the contribution of the phase-field to the reversible stress can only be omitted as long as the phase-field and the density are decoupled. The problem is naturally resolved in the quasi-incompressible hydrodynamics, where $\phi(z)=\phi_0(z)$, $\rho(z)=1$ and $v(z)=0$ indeed represent propagating interface solution. Since $p'[\phi_0(z)]/\phi_0'(z) \equiv -3$, the front speed reads:
\begin{equation}
V = -3\,\kappa_0\,\lambda \enskip ,
\end{equation} 
while the corresponding pressure correction reads $\delta p(z) = \frac{\lambda}{4} \left[\tanh ^3(z)-3 \tanh (z)+2\right]$. The results show that the quasi-incompressible framework recovers the constant density - zero velocity propagating interface in the gapless limit, without the need for altering the right-hand side of the Navier-Stokes equation. Finally, the dispersion relation around $\phi=0,1$ and $1/2$ read:
\begin{eqnarray}
\omega_\phi(k) &=& -3\,\kappa_0(4+k^2) \quad \text{for} \quad \phi=0,1 \enskip ;\\
\label{Model0dispI} \omega_\phi(k) &=& 3\,\kappa_0(2-k^2) \quad \text{for} \quad \phi=1/2 \enskip ,
\end{eqnarray}
which indicate that the bulk solid and liquid are always stable, while the intermediate homogeneous solution is unstable for $|k|<\sqrt{2}$, and therefore the only stable planar interface in the quasi-incompressible dynamics is the solid-liquid interface.

\subsection{Model 1}

The standard method of incorporating the solid-liquid density contrast in the phase-field model is weighting the density dependent free energy densities of the bulk phases locally. The simplest density-phase coupling reads:
\begin{equation}
\label{Model1}
f_\rho(\phi,\rho) := p(\phi) f_s(\rho) + [1-p(\phi)] f_l(\rho) \enskip , 
\end{equation}
where $f_{s,l}(\rho) := [\rho-(1\pm\epsilon)]^2/2$ [see Fig 2(b)]. Eq. (\ref{Model1}) yields:
\begin{eqnarray}
\label{derphimod1}\partial_\phi f_\rho(\phi,\rho) &=& 6 \,\phi\,(1-\phi)\,[2\,\epsilon\,(1-\rho)] \\
\label{derrhomod1}\partial_\rho f_\rho(\phi,\rho) &=& \rho-\{ 1 + \epsilon\,[ 2\,p(\phi)-1]\}  \enskip .
\end{eqnarray}
Using Equations (\ref{derphimod1}) and (\ref{derrhomod1}) in Equations (\ref{planar1}) and (\ref{planar2}) yields the homogeneous equilibrium solutions $\phi=0,1$ and $\phi=[1+B\epsilon(1-\rho)]/2$. To find the solid-liquid interface, we solve the common tangent construction for Eq. (\ref{Model1}), which yields $\rho_{s,l}=1\pm\epsilon$ with $\mu_0=0$. Using Eq. (\ref{Model1}) and $\mu_0=0$ in Eq. (\ref{planar2}) results in $\rho_0(x) = 1+\epsilon\{2\,p[\phi_0(x)]-1\}$, which indicates
\begin{equation}
\label{model1dens}
h(\phi)=1+\epsilon[2\,p(\phi)+1] \enskip .
\end{equation}
Plugging Equations (\ref{Model1}) and (\ref{model1dens}) into Eq. (\ref{frontVsimple}) results in $\sqrt{\frac{3}{2}}\lim_{\phi\to 0,1}\frac{p'(\phi)}{\sqrt{S(q)}} = \lim_{\phi\to 0,1} \frac{3 \sqrt{3}}{\sqrt{3+B \epsilon ^2 [3+4 \phi(1-\phi)]}}$, which shows, that the equilibrium planar interface solution is not an exact solution of the quasi-incompressible dynamics. {Eq. (\ref{frontV}) results in the front velocity}
\begin{equation}
\label{model1V}
V = - \frac{3\kappa_0\,\lambda}{1+\epsilon}\, \frac{1}{\sqrt{1+B\epsilon^2}}  \enskip ,
\end{equation}
{which} suggests, that while the solidification front is "pushed" by the liquid flowing towards the front for $\epsilon>0$ (shrinkage), thus resulting in reduction in the front speed compared to the gapless case, the front is "pulled" for $\epsilon<0$ (expansion), when the liquid moves in the same direction as the front, and the corresponding front velocity is higher than that of the gapless front. {This result is in excellent agreement with previous theoretical predictions.} The dispersion relation around $\phi=0,1$ read:
\begin{equation}
\label{Model1dispphi}
\omega_\phi(k) = -3\,\kappa_0\frac{4(1+B\epsilon^2+k^2)}{1\mp \epsilon} \enskip ,
\end{equation}
thus indicating that the bulk solid and liquid phases are stable for $|\epsilon| < 1$. The intermediate homogeneous solution of the model reads $\phi=[1+B\epsilon(1-\rho)]/2$, which, combined with Eq. (\ref{model1dens}) results in the following solutions: $\phi=1/2$ and $\phi_{1,2}=\frac{1}{2} \left(1\pm\frac{\sqrt{B \epsilon ^2 \left(3 B \epsilon ^2+2\right)}}{B \epsilon ^2}\right)$. The corresponding dispersion relations reads:
\begin{eqnarray}
\omega_\phi(k) &=& 3\,\kappa_0(2+3\,B\epsilon^2-k^2) \enskip ; \\
\omega_\phi(0) &=&  3 B \kappa_0 \frac{  B \epsilon ^2 \left(12 B \epsilon ^2-k^2+20\right)+8}{B^2 \epsilon ^2\mp\sqrt{B \left(3 B \epsilon ^2+2\right)}} \enskip ,
\end{eqnarray}
respectively. The results show that the $\phi=1/2$ homogeneous solution is unstable for $|k|<\sqrt{2+3\,B\epsilon^2}$, while the $\phi_{1,2}$ solutions are also unstable for $|k|<\frac{2}{{\sqrt{B} \epsilon }} \sqrt{\left(B \epsilon ^2+1\right) \left(3 B \epsilon ^2+2\right)}$. These indicate that the only stable interface in the dynamics is the solid-liquid interface. Finally we investigate the properties of the planar equilibrium planar interface. Plugging Equations (\ref{Model1}) and (\ref{model1dens}) into Eq. (\ref{planar1}) results in a second order autonomous ordinary differential equation, which is analytically solvable for $\phi_0(x)$. The solution reads (for details, see Appendix E):
\begin{equation}
\label{model1profile}
\phi_0(x) = \frac{3 \beta+\tanh(y) \{ \alpha \tanh(y)- \xi [1-\tanh(y)] \} }{6 \beta +2 \alpha  \tanh ^2\left(y\right)} \enskip ,
\end{equation}
where $\xi = \frac{1}{2}\sqrt{2\gamma e^{2 y} \left[3+2 \alpha +\gamma \cosh \left(2 y\right)\right]}$, $\alpha=B\epsilon^2$, $\beta=1+\alpha$, $y=\sqrt{\beta}x$, and $\gamma=4 \alpha +3$. Eq. (\ref{model1profile}) and Eq. (\ref{basicprofile}) are compared in Fig 3(a) for different $\epsilon$ values at $B=100$.
\begin{figure}
\includegraphics[width=0.85\linewidth]{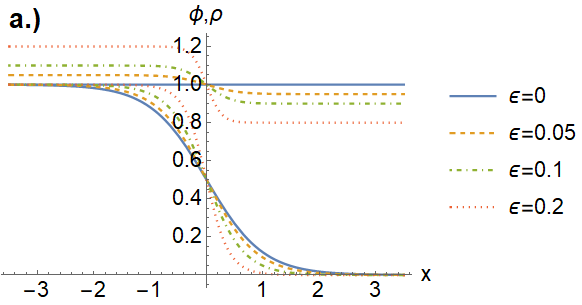}\\
\vspace{0.2cm}
\includegraphics[width=0.85\linewidth]{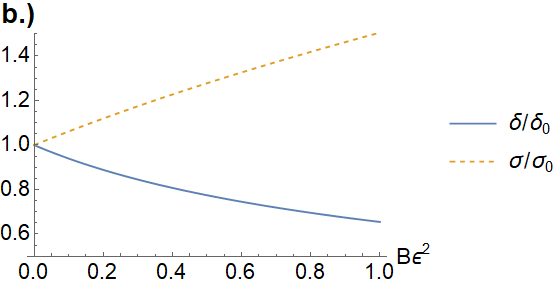}
\caption{Equilibrium planar solid-liquid interfaces and their properties in Model 1. a.) Phase-field (varying between $0$ and $1$, {corresponding to the bulk liquid and solid phases, respectively}) and density (varying between $1-\epsilon$ and $1+\epsilon$, {corresponding to the bulk liquid and solid densities, respectively}) profiles for $\epsilon = 0,0.05,0.1$, and $0.2$ at $B=100$. b.) Dimensionless interface thickness ($\delta/\delta_0$) and interfacial tension ($\sigma/\sigma_0$) as a function of $\alpha=B\epsilon^2$. Note the sharpening interface (characterized by decreasing interface thickness) and increasing interfacial tension with increasing density gap at constant bulk modulus.}
\end{figure}
The characteristic interface width and the solid-liquid interfacial tension read:
\begin{eqnarray}
\label{deltamodel1}f_\delta(\alpha) &:=& -\frac{1}{2 \phi'_0(x)|_{x=0}} = \sqrt{3/\gamma} \enskip ; \\
\label{sigmamodel1}f_\sigma(\alpha) &=& \frac{9\left[ \sqrt{\alpha  \beta} (\alpha+\beta) - \frac{\gamma}{\sqrt{3}} \cot ^{-1}\left(\sqrt{\frac{3}{\alpha }+3}\right) \right]}{16 \alpha ^{3/2}} \enskip . \quad \quad
\end{eqnarray}
Note that Equations (\ref{model1profile})-(\ref{sigmamodel1}) recover $\phi_0(x)=[1-\tanh(x)]/2$ and $f_\delta=f_\sigma=1$ in the gapless ($\alpha\to 0$) limit. For $\alpha\neq 0$ the equations indicate that the density change alters the shape and the width of the equilibrium planar interface, and also contributes to its energy. The dimensional interface width and interfacial deviate from $\delta$ and $\sigma$ by approximately $15\%$ for $\alpha = 100 \times 0.05^2 = 1/4$ [see Fig. 3(b)],  and therefore these parameters must-be re-calculated. Denoting the dimensional interface width and interfacial free energy by $\tilde{\delta}$ and $\tilde{\sigma}$, respectively, the following implicit equations must be solved for $\delta$ and $\sigma$ at fixed $\tilde{\alpha}:=K\epsilon^2$:
\begin{equation}
\tilde{\delta} = \delta f_\delta\left(\tilde{\alpha}\,\delta/\sigma\right) \quad \text{and} \quad \tilde{\sigma} = \sigma f_\sigma\left(\tilde{\alpha}\,\delta/\sigma\right) \enskip ,
\end{equation}
which can be solved numerically.

\subsection{Model 2}

\begin{figure}
\includegraphics[width=0.85\linewidth]{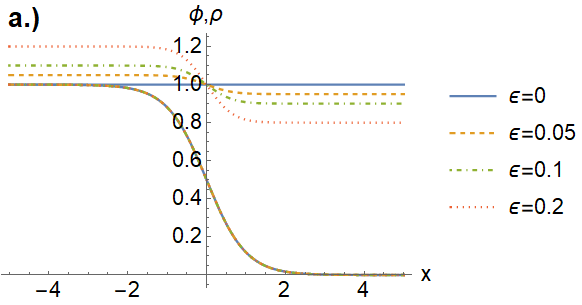}\\
\vspace{0.2cm}
\includegraphics[width=0.85\linewidth]{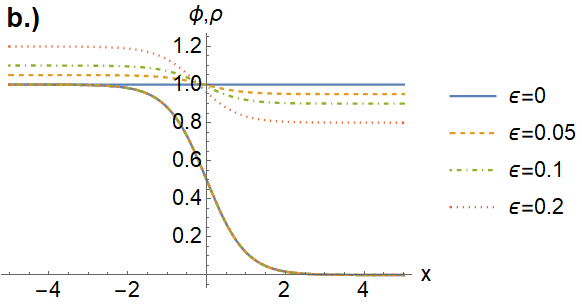}
\caption{Equilibrium planar solid-liquid interfaces for Model 2 for a.) $q(\phi)=1+\epsilon[2\,p(\phi)-1]$ and b.) $q(\phi)=1/[\phi/(1+\epsilon)+(1-\phi)/(1-\epsilon)]$. The density profiles vary between $1-\epsilon$ and $1+\epsilon$. Note that the phase-field profile (varying between $0$ and $1$, {corresponding to the bulk liquid and solid phases, respectively}) is independent of $q(\phi)$.}
\end{figure}

To avoid the difficulties related to fitting the model parameters $\sigma$ and $\delta$ in Model 1, and to guarantee that the equilibrium planar interface represents exact solution for the propagating interface in the quasi-incompressible model, we propose the following density-phase coupling:
\begin{equation}
\label{Model2}
f_\rho(\phi,\rho) := \frac{1}{2}\left[ \rho-q(\phi) \right]^2 \enskip , 
\end{equation}
where we prescribe the conditions $q(0):=1-\epsilon$ and $q(1):=1+\epsilon$ for $q(\phi)$ [see Fig 2(c) and (d)]. Using Eq. (\ref{Model2}) in Eq. (\ref{Model1}) yields:
\begin{eqnarray}
\label{derphimod2}\partial_\phi f_\rho(\phi,\rho) &=& [\rho-q(\phi)]q'(\phi) \\
\label{derrhomod2}\partial_\rho f_\rho(\phi,\rho) &=& [\rho-q(\phi)]  \enskip .
\end{eqnarray}
The homogeneous solutions of the Euler-Lagrange equations can be determined as follows. The Euler-Lagrange equation $\delta_\rho F = B \partial_\rho f_\rho = B[\rho-q(\phi)]=\mu$ indicates $\delta_\phi F = \delta_\phi F_0+B [\rho-q(\phi)]q'(\phi)=\delta_\phi F_0 + \mu\,q'(\phi)=0$. The solutions of the latter are $\phi=0,1$ [for at least $\rho=1\mp\epsilon$, respectively], while further solutions may also be possible, depending on the particular form of $q(\phi)$. The common tangent construction results in the coexistence densities $\rho_{s,l}=1\pm\epsilon$ and $\mu_0=0$ again for arbitrary $q(\phi)$ satisfying the conditions $q(0):=1-\epsilon$ and $q(1):=1+\epsilon$, which, when used in Eq. (\ref{planar2}), indicates $\rho_0(x) = q[\phi_0(x)]$, which yields
\begin{equation}
\label{model2dens}
h(\phi) \equiv q(\phi) \enskip .
\end{equation}
Moreover, $\phi_0(x)=[1-\tanh(x)]/2$ independently of $q(\phi)$ and $B$ (see Fig 4)! Using Equations (\ref{Model2}) and (\ref{model2dens}), together with $S(\phi) \equiv g(\phi)$ in Eq. (\ref{frontVsimple}) results in $\sqrt{\frac{3}{2}}\,\frac{p'(\phi)}{\sqrt{S(q)}} \equiv 3$, which indicates that the planar equilibrium interface is an \textit{exact} solution of the quasi-incompressible hydrodynamics with
\begin{equation}
\label{model2V}
V = - \frac{3\kappa_0\,\lambda}{1+\epsilon} \enskip .
\end{equation}
The corresponding phase-field, density, and velocity profiles are shown in Fig 5. for negative and positive relative density gap, thus illustrating decelerated ("pushed") and accelerated ("pulled") solidification fronts, respectively. A further consequence of Eq. (\ref{Model2}) is that the density change doesn't contribute to the interfacial free energy of the equilibrium planar interface, and therefore $\delta$ and $\sigma$ coincide with the interface width and the interfacial tension (respectively) for arbitrary $q(\phi)$, and no implicit equations need to be solved for them. A further simplification is that the condition of quasi-incompressibility, $\rho(\mathbf{r},t):=q[\phi(\mathbf{r},t)]$ results in
\begin{equation}
\label{condmodel2}
\delta_\phi^{(c)} F \equiv \delta_\phi F_0
\end{equation}
in Equations (\ref{qeq1}) and (\ref{qeq3}), and therefore the equation of state $\rho=q(\phi)$ cancels the functional derivative. The dispersion relation around $\phi=0,1$ coincide with Eq. (\ref{Model1dispphi}), and therefore the bulk solid and liquid phases are stable. To test the stability of the equilibrium planar interface, we choose
\begin{equation}
q(\phi):=1+\epsilon[2\,p(\phi)+1] \enskip ,
\end{equation}
which provides the intermediate homogeneous solution $\phi=\frac{1}{2}(1+\epsilon\mu)$ for arbitrary chemical potential. Since $\mu_0=0$ for the solid-liquid planar equilibrium interface, the only intermediate homogeneous equilibrium solution in the quasi-incompressible dynamics $\phi=1/2$, for which the dispersion relation coincides with Eq. (\ref{Model0dispI}), showing that the only stable interface in the model is, again, the solid-liquid interface. Following Kim and Lowengrub \cite{KimLowengrub2005}, another reasonable choice can be
\begin{equation}
\label{Model2test}
q(\phi):=\frac{1}{\frac{\phi}{1+\epsilon}+\frac{1-\phi}{1-\epsilon}} \enskip ,
\end{equation}
which has a numerical advantage, since $s:=(dq/d\phi)/q^2(\phi)=2\,\epsilon/(1-\epsilon^2)$ (constant) in this case, for which Eq. (\ref{qeq3}) reads $\nabla\cdot\mathbf{v}=s\,\kappa_0\,\left( \delta_\phi F_0 \right)$, where we used Eq. (\ref{condmodel2}). Finally, the dispersion relations around $\phi=0,1$ read: $\omega_\phi(k)=-3\,\kappa_0\frac{4+k^2}{1\mp \epsilon}$, while the only extra homogeneous phase is $\phi=1/2$, for which the dispersion relation reads: $\omega_\phi(k) = 3\,\kappa_0 \,\frac{2-k^2}{1-\epsilon^2}$. These also indicate that the intermediate bulk phase is unstable, and that the only stable interface in the quasi-incompressible dynamics is the equilibrium solid-liquid interface.  
\begin{figure}
\includegraphics[width=0.85\linewidth]{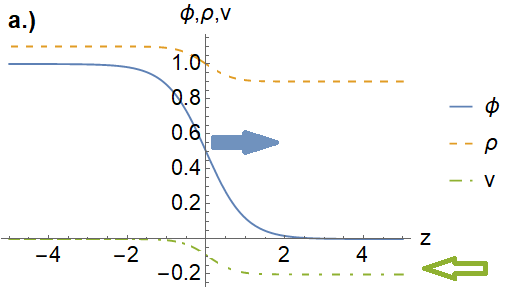}\\
\vspace{0.2cm}
\includegraphics[width=0.85\linewidth]{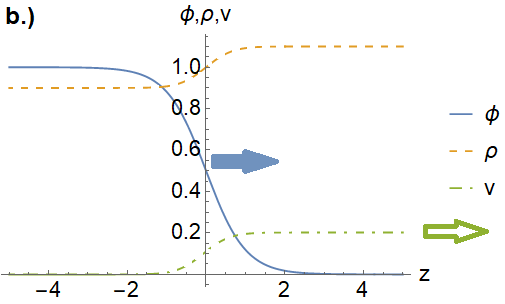}
\caption{Exact propagating planar solid-liquid interface {(with bulk liquid and solid phases at $z \to \pm\infty$, respectively)} solutions of the quasi-incompressible hydrodynamic equations in Model 2 for $q(\phi)=1+\epsilon[2\,p(\phi)-1]$, $\epsilon=+0.1$ [panel a.)], and $\epsilon=-0.1$ [panel b.)]. The velocity profile is normalized by $-3\,\lambda\,\kappa_0 > 0$ (for $\lambda<0$). The arrows indicate the direction of the solidification front {(bulk)} and the fluid flow {(empty)}. Note that $\epsilon > 0$ indicates that the fluid moves against the front ("pushed" front), while the front and the fluid move in the same direction in case of $\epsilon<0$ ("pulled" front).}
\end{figure} 

\section{Summary}

In this work we developed a physically consistent, analytically easily calibratable, and numerically efficient hydrodynamic phase-field model incorporating density change accompanying the structural transition. The derivation of the general hydrodynamic equations started from the balance equations for the microscopic liquid and solid mass and momentum densities. To obtain coarse-grained equations being valid on the length scale of the solid-liquid interface width, the statistical physical ensemble averages of the microscopic phase-change rate, advective and capillary stresses were approximated by using standard statistical physical methods (ideal gas approximation and adiabatic approximation) and the principles of non-equilibrium thermodynamics (constitutive relation for the phase-change rate on the variational basis). Furthermore, we introduced the principle of co-moving phases, which expresses, that the solid-liquid transition is a local structural change, and therefore the phases cannot move relative to each other (i.e., by diffusion). In the next step of the model development we have shown, that the general form of the right-hand side of the coarse-grained Navier-Stokes equation is limited by momentum conservation. Since the total coarse-grained momentum of the system coincides with the total microscopic momentum, only such approximations of the ensemble average of the microscopic advective and capillary stresses are valid, for which the total coarse-grained momentum is conserved in non-equilibrium. This principle is satisfied, as long as such an $\mathbb{R}$ tensor exists (being the function of the coarse-grained variables and their spatial derivatives), for which $\nabla\cdot\mathbb{R}$ coincides with the right-hand side of the coarse-grained momentum equation. This resulted the Navier-Stokes equation
\begin{equation*}
\partial_t (\rho\,\mathbf{v})+\nabla\cdot(\rho\,\mathbf{v}\otimes\mathbf{v}) = -\rho\nabla(\delta_\rho F)+(\delta_\phi F)\nabla\phi
\end{equation*} 
for $\phi$ being the local mass fraction of the solid. This result is in excellent agreement with the works of Anderson, Wheeler and McFadden and Conti, but differs from the {equation used by} Oxtoby and Harrowell. Postulating the right-hand side of the Navier-Stokes equation was followed by the construction of the free energy functional, which has been done on the basis of the standard Phase-Field Theory. Using linear stability analysis we have shown, that the inviscid compressible hydrodynamic equations contain 2 time scales. One is associated with the relaxation of the phase, whereas the other is the time scale of sound waves. In order to make the time scale of pattern formation accessible in numerical simulations, we introduced quasi-incompressible hydrodynamic equations [see Equations (\ref{qeq1})-(\ref{qeq3})], in which sound waves are eliminated, even though the density can vary across the solid-liquid interface. In the next step, both the compressible and the quasi-incompressible dynamics' were investigated against equilibrium and propagating planar interfaces in general. It has been found, that a dynamical expansion of the solid is present in the compressible model even in the gapless limit (when the equilibrium solid-liquid density gap is zero). {This artificial phenomenon is absent in the quasi-incompressible dynamics, where the front velocity can be calculated for arbitrary density gap without determining the propagating interface. The calculation yielded
\begin{equation}
\label{central}
V = -\frac{3\,\Gamma_0\,\Lambda\,\delta}{\bar{\rho}(1+\epsilon)}[1+O(\epsilon^2)]
\end{equation}
for the dimensional front velocity, where $\Gamma_0$ is a constant phase-field mobility, $\Lambda$ the thermodynamic driving force, $\bar{\rho}=\rho_s^0+\rho_l^0$ the average coexistence density, $\rho_{l,s}^0$ the coexistence liquid and solid densities, respectively, and $\epsilon=(\rho_s-\rho_l)/(\rho_s+\rho_l)$ the half of the relative density gap.} Eq. (\ref{central}) indicates the presence of the so-called "pushed" and "pulled" solidification fronts: When the forming solid is denser than the liquid, more liquid needs to undergo the structural transition in the interface region, and therefore the front speed is reduced compared to a gapless system. In contrast, for negative relative density contrast, when the solid expands upon solidification, the fluid flow removes excess liquid from the interface region, and therefore the front is faster compared to the gapless case. These findings are also in excellent agreement with the results of Tien and Koump, {Oxtoby and Harowell} (see Appendix F), and Conti. Finally, we tested the calibratability of the model for two different density-phase couplings. For Model 1 [described by Eq. (\ref{Model1})] it was found, that the equilibrium interface was a complicated function of the model parameters, which needed to be fitted numerically. In contrast, a simplified density-phase coupling (called Model 2) recovered the equilibrium planar interface $\phi_0(x)=[1-\tanh(x)]/2$ and $\rho_0(x)=q[\phi_0(x)]$ for arbitrary equation of state $\rho=q(\phi)$. Moreover, since the density change doesn't contribute to the equilibrium interface, the model parameters don't need to be fitted numerically. Finally, the equilibrium planar interface solution always represent exact propagating planar interface solution of the quasi-incompressible hydrodynamics for arbitrary equation of state, with front velocity $V = -\frac{3\,\lambda\,\kappa_0}{1+\epsilon}$ for constant phase-field mobility. These properties make the quasi-incompressible model described by Equations (\ref{qeq1})-(\ref{qeq3}), when combined with Model 2, an easily calibratable, and numerically efficient modelling tool to study pattern formation in the presence of solidification shrinkage/expansion. In the future, we plan to extend the quasi-incompressible framework for binary materials to study the effect of density change on dendrite growth.

\section*{Acknowledgements}

The authors thank Professor Mathis Plapp (\'Ecole Polytechnique, Paris, France), {Professor L\'aszl\'o Gr\'an\'asy (Wigner Research Centre for Physics, Budapest, Hungary)}, Dr. Ricardo L. Barros (Loughborough University, Loughborough, UK), {and Frigyes Podmanizcky (Wigner Research Centre for Physics, Budapest, Hungary)} for their support and valuable comments. 

\section*{APPENDIX}

\subsection*{A. Fundamental dynamical equations and coarse-graining}

Consider a system of $N$, pair interacting classical particles. The Hamiltonian of the system reads:
\begin{equation}
\label{AppAH}
\hat{H} = \sum_{i}\frac{\mathbf{p}_i(t)^2}{2\,m} + \hat{V} \enskip ,
\end{equation} 
where $m$ is the particle mass, $\hat{V}=\frac{1}{2}\sum_{i,j}v(|\mathbf{r}_i(t)-\mathbf{r}_j(t)|)$ the potential energy, and $\mathbf{v}(r)$ the pair potential.  The time evolution of the system is driven by the canonical equations:
\begin{eqnarray*}
\dot{\mathbf{r}}_i(t) &=& \mathbf{p}_i(t)/m \enskip ;\\
\dot{\mathbf{p}}_i(t) &=& -\partial_{\mathbf{r}_i(t)} \hat{V}  \enskip .
\end{eqnarray*}
The microscopic solid and liquid mass and momentum densities are defined as follows:
\begin{eqnarray*}
\hat{\rho}_s(\mathbf{r},t) &:=& m \sum_i \phi_i(t) \delta[\mathbf{r}-\mathbf{r}_i(t)] \enskip ; \\
\hat{\rho}_l(\mathbf{r},t) &:=& m \sum_i [1-\phi_i(t)] \delta[\mathbf{r}-\mathbf{r}_i(t)] \enskip ;\\
\hat{\mathbf{g}}_s(\mathbf{r},t) &:=& \sum_i \phi_i(t) \mathbf{p}_i(t) \delta[\mathbf{r}-\mathbf{r}_i(t)] \enskip ; \\
\hat{\mathbf{g}}_l(\mathbf{r},t) &:=& \sum_i [1-\phi_i(t)] \mathbf{p}_i(t) \delta[\mathbf{r}-\mathbf{r}_i(t)] \enskip ,
\end{eqnarray*}
respectively, where $\phi_i(t) \in [0,1]$ is the normalized bond order parameter of particle $i$ ($0$ and $1$ denote the bulk liquid and solid phases, respectively). The time evolution of the microscopic densities can be calculated as follows. First we define the forward and inverse Fourier transform of a quantity as $\hat{Y}(\mathbf{k},t):=\frac{1}{(2\pi)^3}\int d\mathbf{r} \left\{ \hat{y}(\mathbf{r},t) e^{-\imath\mathbf{k}\cdot\mathbf{r}} \right\}$ and $\hat{y}(\mathbf{r},t)=\int d\mathbf{k} \left\{ \hat{Y}(\mathbf{k},t) e^{\imath\mathbf{k}\cdot\mathbf{r}} \right\}$, respectively. Using these definitions, the Fourier transform of the solid mass and momentum densities read:
\begin{eqnarray*}
\hat{P}_s(\mathbf{k},t) &=& \frac{m}{(2\pi)^3}\sum_i \phi_i(t) e^{-\imath\mathbf{k}\cdot\mathbf{r}_i(t)} \enskip ; \\
\hat{\mathbf{G}}_s(\mathbf{k},t) &=& \frac{1}{(2\pi)^3}\sum_i \phi_i(t) \mathbf{p}_i(t) e^{-\imath\mathbf{k}\cdot\mathbf{r}_i(t)} \enskip .
\end{eqnarray*}
Differentiating $\hat{P}_s(\mathbf{k},t)$ with respect to time yields:
\begin{equation*}
\begin{split}
\partial_t \hat{P}_s =& \frac{m}{(2\pi)^3}\sum_i (\dot{\phi}_i - \imath\mathbf{k} \cdot \dot{\mathbf{r}}_i\phi_i) e^{-\imath\mathbf{k}\cdot\mathbf{r}_i} = -\imath\mathbf{k} \cdot \hat{\mathbf{G}}_s + \hat{\Sigma} \enskip ,
\end{split}
\end{equation*}
where $\hat{\Sigma}(\mathbf{k},t) = \frac{m}{(2\pi)^3}\sum_i \dot{\phi}_i(t) e^{-\imath\mathbf{k}\cdot\mathbf{r}_i(t)}$. Re-arranging the equation and inverse Fourier transforming it results in:
\begin{equation}
\label{AppArhoS}
\partial_t \hat{\rho}_s+\nabla\cdot\hat{\mathbf{g}}_s=\hat{\sigma} \enskip ,
\end{equation}
where $\hat{\sigma}(\mathbf{r},t) = m \sum_i \dot{\phi}_i(t) \delta[\mathbf{r}-\mathbf{r}_i(t)]$. Repeating the steps of the derivation for the liquid mass density yields a similar equation:
\begin{equation}
\label{AppArhoL}
\partial_t \hat{\rho}_l+\nabla\cdot\hat{\mathbf{g}}_l=-\hat{\sigma} \enskip ,
\end{equation}
while the adding Equations (\ref{AppArhoS}) and (\ref{AppArhoL}) results in microscopic continuity:
\begin{equation}
\label{AppAcont}
\partial_t \hat{\rho}+\nabla\cdot\hat{\mathbf{g}}=0 \enskip ,
\end{equation}
where $\hat{\rho}(\mathbf{r},t)=\hat{\rho}_s(\mathbf{r},t)+\hat{\rho}_l(\mathbf{r},t)$ and $\hat{\mathbf{g}}(\mathbf{r},t)=\hat{\mathbf{g}}_s(\mathbf{r},t)+\hat{\mathbf{g}}_l(\mathbf{r},t)$. The solid and liquid momentum equations are derived similarly. First we write the pair potential in the form $v(\mathbf{r}) = \int d\mathbf{k} \left\{ V(\mathbf{k})e^{\imath\mathbf{k}\cdot\mathbf{r}} \right\}$, where $V(\mathbf{k}) = \frac{1}{(2\pi)^3}\int d\mathbf{r} \left\{ v(\mathbf{r})e^{-\imath\mathbf{k}\cdot\mathbf{r}} \right\}$ is the Fourier transform of the pair potential \cite{Zaccarelli_2002}. Differentiating the solid momentum density with respect to time yields:
\begin{equation*}
\begin{split}
\partial_t \hat{\mathbf{G}}_s = & \frac{1}{(2\pi)^3}\sum_i ( \dot{\mathbf{p}}_i\phi_i+ \mathbf{p}_i\dot{\phi}_i -\imath\mathbf{k}\cdot\dot{\mathbf{r}}_i\phi_i\mathbf{p}_i ) e^{-\imath\mathbf{k}\cdot\mathbf{r}_i}  = \\
& - \frac{1}{(2\pi)^3}\sum_i (\partial_{\mathbf{r}_i} \hat{V})\phi_ie^{-\imath\mathbf{k}\cdot\mathbf{r}_i} -\imath\mathbf{k} \cdot \hat{\mathbf{K}}_s + \hat{\mathbf{J}} \enskip , 
\end{split}
\end{equation*}
where $\hat{\mathbf{K}}_s(\mathbf{k},t)=\frac{1}{(2\pi)^3}\sum_i \phi_i(t)\frac{\mathbf{p}_i(t)\otimes\mathbf{p}_i(t)}{m} e^{-\imath\mathbf{k}\cdot\mathbf{r}_i(t)}$ and $\hat{\mathbf{J}}(\mathbf{k},t) = \frac{1}{(2\pi)^3}\sum_i \dot{\phi}_i(t) \mathbf{p}_i(t) e^{-\imath\mathbf{k}\cdot\mathbf{r}_i(t)}$. Using $\hat{V}=\frac{1}{2}\sum_{i,j}v(\mathbf{r}_i-\mathbf{r}_j) = \frac{1}{2}\sum_{i,j} \int d\mathbf{k} \left\{ V(\mathbf{k})e^{\imath\mathbf{k}\cdot(\mathbf{r}_i-\mathbf{r}_j)} \right\}$ result in:
\begin{equation*}
\partial_{\mathbf{r}_i} \hat{V} = \sum_j \int d\mathbf{q} \left\{ \imath\mathbf{q}V(\mathbf{q})e^{\imath\mathbf{q}\cdot[\mathbf{r}_i(t)-\mathbf{r}_j(t)]} \right\} \enskip ,
\end{equation*}
which can be used in the momentum equation, yielding:
\begin{equation*}
\begin{split}
\partial_t \hat{\mathbf{G}}_s = & (2\pi)^3 \int d\mathbf{q} \left\{ -\imath\mathbf{q}\frac{V(\mathbf{q})}{m^2} \hat{P}(\mathbf{q},t) \hat{P}_s(\mathbf{k}-\mathbf{q},t) \right\} \\
& -\imath\mathbf{k} \cdot \hat{\mathbf{K}}_s + \hat{\mathbf{J}} \enskip , 
\end{split}
\end{equation*}
where $\hat{P}(\mathbf{q},t)=\hat{P}_s(\mathbf{q},t)+\hat{P}_l(\mathbf{q},t)$. After inverse Fourier transform we get:
\begin{equation}
\label{AppAmomS}
\partial_t \hat{\mathbf{g}}_s+\nabla\cdot\hat{\mathbb{K}}_s=-\hat{\rho}_s \nabla\left(\frac{v}{m^2} * \hat{\rho}\right) + \hat{\mathbf{j}} \enskip ,
\end{equation}
where $*$ stands for spatial convolution, while the advective stress tensor and the momentum current read: 
\begin{eqnarray*}
\hat{\mathbb{K}}_s(\mathbf{r},t)&=&\sum_i \phi_i(t) \frac{\mathbf{p}_i(t) \otimes \mathbf{p}_i(t)}{m} \delta[\mathbf{r}-\mathbf{r}_i(t)] \enskip ;\\
\hat{\mathbf{j}}(\mathbf{r},t)&=&\sum_i \dot{\phi}_i(t) \mathbf{p}_i(t) \delta[\mathbf{r}-\mathbf{r}_i(t)] \enskip .
\end{eqnarray*}
A similar derivation can be carried out for the liquid momentum density, yielding:
\begin{equation}
\label{AppAmomL}
\partial_t \hat{\mathbf{g}}_l+\nabla\cdot\hat{\mathbb{K}}_l=-\hat{\rho}_l \nabla\left(\frac{v}{m^2} * \hat{\rho}\right) - \hat{\mathbf{j}} \enskip ,
\end{equation}
where $\hat{\mathbb{K}}_l(\mathbf{r},t)=\sum_i [1-\phi_i(t)] \frac{\mathbf{p}_i(t) \otimes \mathbf{p}_i(t)}{m} \delta[\mathbf{r}-\mathbf{r}_i(t)]$. The sum of the two momentum equations results in the microscopic Navier-Stokes equation:
\begin{equation}
\label{AppAmom}
\partial_t \hat{\mathbf{g}}+\nabla\cdot\hat{\mathbb{K}}=-\hat{\rho} \nabla\left(\frac{v}{m^2} * \hat{\rho}\right) \enskip ,
\end{equation}
where $\hat{\mathbb{K}}(\mathbf{r},t)=\hat{\mathbb{K}}_s(\mathbf{r},t)+\hat{\mathbb{K}}_l(\mathbf{r},t)$. Using the convolution theorem $(f * g)(x)=\int dx' f(x')g(x-x')=\int dx' g(x')f(x-x')$, the Fourier transform of the equation at $\mathbf{k}=\mathbf{0}$ reads:
\begin{equation*}
\partial_t\hat{\mathbf{G}}(\mathbf{0},t) =- \frac{(2\pi)^3}{2\,m^2}\int d\mathbf{q} \, \imath\mathbf{q}\left[V(\mathbf{q})-V(-\mathbf{q})\right] ||\hat{P}(\mathbf{q},t)||^2 \enskip .
\end{equation*}
Since $(2\pi)^3\hat{\mathbf{G}}(\mathbf{0},t)=\int d\mathbf{r}\,\hat{\mathbf{g}}(\mathbf{r},t)=\sum_i \mathbf{p}_i(t)$ is the full momentum, and $v(\mathbf{r})=v(-\mathbf{r})$ indicates $V(\mathbf{k})=V(-\mathbf{k})$ (Newton's third law), the above equation reads:
\begin{equation}
\label{AppAmomC}
\frac{d}{dt} \int d\mathbf{r} \, \hat{g}(\mathbf{r},t) = \mathbf{0} \enskip ,
\end{equation}
i.e., the full momentum is conserved in Eq. (\ref{AppAmom}). The right-hand side of Eq. (\ref{AppAmom}) can also be expressed in terms of the potential energy. Using the Fourier representation of the pair potential and the total mass density results in:
\begin{equation*}
\begin{split}
\hat{V} = & \frac{1}{2}\sum_{i,j} v[\mathbf{r}_i(t)-\mathbf{r}_j(t)] = \frac{1}{2}\sum_{i,j}\int d\mathbf{k} V(\mathbf{k}) e^{\imath\mathbf{k}(\mathbf{r}_i-\mathbf{r}_j)} = \\
& \frac{(2\pi)^6}{2\,m^2} \int d\mathbf{k} \, V(\mathbf{k}) \hat{P}(\mathbf{k},t) \hat{P}(-\mathbf{k},t) = \\
& \frac{1}{2} \int d\mathbf{r}\, \left\{ \hat{\rho}(\mathbf{r},t) \int d\mathbf{r}' \, \hat{\rho}(\mathbf{r}',t) \int d\mathbf{r}'' \, \frac{v(\mathbf{r}'')}{m^2} \right. \\ 
& \left. \times \frac{1}{(2\pi)^3} \int d\mathbf{k}\,e^{\imath\mathbf{k}(\mathbf{r}'-\mathbf{r}-\mathbf{r}'')} \right\} = \frac{1}{2} \int d\mathbf{r} \left\{\hat{\rho} \left(\frac{v}{m^2} * \hat{\rho} \right) \right\} \enskip .
\end{split}
\end{equation*} 
Since $v(\mathbf{r})=v(-\mathbf{r})$, the first functional derivative of $\hat{V}$ with respect to $\hat{\rho}$ reads:
\begin{equation*}
\delta_{\hat{\rho}} \hat{V} = \frac{v}{m^2} * \hat{\rho} \enskip .
\end{equation*}
Using this expression in Eq. (\ref{AppAmom}) yields:
\begin{equation}
\label{AppAmomF}
\partial_t \hat{\mathbf{g}}+\nabla\cdot\hat{\mathbb{K}}=-\hat{\rho}\nabla(\delta_{\hat{\rho}}\hat{V}) \enskip ,
\end{equation}
which already implies the variational form of the right-hand side of the coarse-grained equation. The macroscopic (or coarse-grained) variables read:
\begin{eqnarray*}
\rho_{s,l}(\mathbf{r},t) &:=& \int d\mathbb{P}(\mathbf{\Gamma}_0)\,\hat{\rho}_{s,l}(\mathbf{r},t) \enskip ;\\
\mathbf{g}_{s,l}(\mathbf{r},t) &:=& \int d\mathbb{P}(\mathbf{\Gamma}_0)\,\hat{\mathbf{g}}_{s,l}(\mathbf{r},t) \enskip ,
\end{eqnarray*}
which is a weighted sum of the trajectories over the distribution of the initial condition. Since the initial condition of the system appears as a set of parameters in the solution of the Hamiltonian system [i.e., formally we can write $\mathbf{r}_i^{\mathbf{\Gamma}_0}(t)$ and  $\mathbf{p}^{\mathbf{\Gamma}_0}_i(t)$], the coarse graining process doesn't operate on $\mathbf{r}$ and $t$. Consequently, coarse-graining Equations (\ref{AppArhoS}), (\ref{AppArhoL}), (\ref{AppAmomS}) and (\ref{AppAmomL}) results in:
\begin{eqnarray*}
\partial_t \rho_{s,l} + \nabla\cdot\mathbf{g}_{s,l} &=& \pm \sigma \\
\partial_t \mathbf{g}_{s,l} + \nabla\cdot\mathbb{K}_{s,l} &=& \pm \mathbf{j}  - \int d\mathbb{P}(\mathbf{\Gamma}_0)\{\hat{\rho}_{s,l}\nabla(\delta_{\hat{\rho}}\hat{V})\} \enskip , \quad
\end{eqnarray*}
where $\sigma(\mathbf{r},t)$, $\mathbf{j}(\mathbf{r},t)$ and $\mathbb{K}_{s,l}(\mathbf{r},t)$ are the coarse-grained variables of their microscopic counterparts. The macroscopic continuity and the Navier-Stokes equation read:
\begin{eqnarray*}
\partial_t \rho + \nabla\cdot\mathbf{g} &=& 0 \\
\partial_t \mathbf{g} + \nabla\cdot\mathbb{K} &=& -\mathbf{f} \enskip , \quad
\end{eqnarray*}
where 
\begin{eqnarray*}
\mathbb{K}(\mathbf{r},t) &=& \int d\mathbb{P}(\mathbf{\Gamma}_0) \sum_i \frac{\mathbf{p}_i(t)\otimes\mathbf{p}_i(t)}{m} \delta[\mathbf{r}-\mathbf{r}_i(t)] \quad \\
\mathbf{f}(\mathbf{r},t) &=& \int d\mathbb{P}(\mathbf{\Gamma}_0)\left\{\hat{\rho}\nabla\sum_i\frac{v[\mathbf{r}_i(t)-\mathbf{r}]}{m}\right\}
\end{eqnarray*} 
are the coarse-grained advective and reversible stress tensors, respectively. For an applicable macroscopic model, we need to approximate $\mathbb{K}$ and $\mathbf{f}$ in terms of the coarse-grained variables in non-equilibrium. The approximation of these terms depend on the length scale of interest and the number of coarse-grained variables (these are included in the free energy functional of the system), but the most important principle in is that $\mathbf{g}$ is also conserved quantity, since:
\begin{equation*}
\frac{d}{dt}\int d\mathbf{r} \, \mathbf{g}(\mathbf{r},t) = \int d\mathbb{P}(\mathbf{\Gamma}_0)\left\{ \frac{d}{dt}\int d\mathbf{r} \, \hat{\mathbf{g}}(\mathbf{r},t) \right\} =0 \enskip ,
\end{equation*}  
which directly emerges from the definition of $\mathbf{g}$ and Eq. (\ref{AppAmomC}). It means that only such approximation of $\mathbf{f}$ is acceptable, for which $\mathbf{F}(\mathbf{0},t)=0$. To investigate what this approximation can be, first we approximate $\mathbb{K}$ by using the ideal gas approximation, yielding:
\begin{equation*}
\mathbb{K} \approx \rho \left[ \mathbf{v}\otimes\mathbf{v}+\left(\frac{k_B T}{m}\right)\,\mathbb{I} \right] \enskip ,
\end{equation*} 
where $\mathbf{v}(\mathbf{r},t)=\mathbf{g}(\mathbf{r},t)/\rho(\mathbf{r},t)$ is the coarse-grained velocity. Putting the second term to the right-hand side of the momentum equation yields:
\begin{equation*}
\partial_t (\rho\mathbf{v}) + \nabla\cdot(\rho\mathbf{v}\otimes\mathbf{v}) = -[\rho\nabla(\delta_\rho F_0)+\mathbf{f}] \enskip , \quad
\end{equation*}
where $F_0=\int d\mathbf{r}\left\{ (k_B T)\frac{\rho}{m}\left[ \log\left(\frac{\rho}{m}\lambda_d^3 \right)-1\right] \right\}$ is the free energy of the ideal gas, and $\lambda_d$ the de-Broglie wavelenght. As long as the only coarse-grained variables are $\rho$ and $\mathbf{g}$, $\mathbf{f}$ can be approximated as $\mathbf{f} \approx \rho\nabla(\delta_\rho F_{ex}[\rho])$ \cite{doi:10.1063/1.4913636}, where $F_{ex}[\rho]$ is the excess free energy of the system (emerging from the presence of interactions), thus yielding:
\begin{equation}
\label{AppAsingle}
\partial_t (\rho\mathbf{v}) + \nabla\cdot(\rho\mathbf{v}\otimes\mathbf{v}) = -\rho\nabla(\delta_\rho F[\rho]) \enskip , \quad
\end{equation}
which becomes $0$ in equilibrium [described by $\delta_\rho F = \mu$ (constant)]. It can be proven that as long as $F[\rho]=\int d\mathbf{r} f(\rho,\nabla\rho,\nabla^2\rho,\dots)$, such an $\mathbb{R}(\rho,\nabla\rho,\nabla^2\rho,\dots)$ tensor exists, for which $\nabla\cdot\mathbb{R} \equiv -\rho \nabla (\delta_\rho F[\rho])$, and therefore Eq. (\ref{AppAsingle}) preserves the total momentum (for a simple example, see Appendix B).\\

Contrary to this simple scenario, the situation can be essentially different for $F[\rho_s,\rho_l]$, or, in general, when the free energy of the system is expressed in terms of multiple coarse-grained variables. In Appendix B we will show, that Eq. (\ref{AppAsingle}) is not suitable to describe systems with coupled coarse-grained variables in $F[.]$. In this case it is better to consider a binary system to find a suitable approximation of $\mathbf{f}$. Replacing the subscripts $s$ and $l$ by $\alpha$ and $\beta$, omitting $\phi_i(t)$ in the mass and momentum densities, and using the Hamiltonian $\hat{H} := \sum_i \frac{[\mathbf{p}^{\alpha}_i(t)]^2}{2 m_\alpha}+\sum_i \frac{[\mathbf{p}^{\beta}_i(t)]^2}{2 m_\beta}+\hat{V}$, where $\hat{V}=\frac{1}{2}\left\{ \sum_{i,j}v_{\alpha\alpha}[\mathbf{r}_i^{\alpha}(t)-\mathbf{r}_j^{\alpha}(t)+\sum_{i,j}v_{\beta\beta}[\mathbf{r}_i^{\beta}(t)-\mathbf{r}_j^{\beta}(t)\right\}+\sum_{i,j}v_{\alpha\beta}[\mathbf{r}_i^{\alpha}(t)-\mathbf{r}_j^{\beta}(t)]$, then repeating the steps of the derivations from Eq. (\ref{AppAH}) result in the following microscopic momentum equations:
\begin{equation*}
\partial_t \hat{\mathbf{g}} + \nabla\cdot\hat{\mathbb{K}} = -\hat{\rho}_{\alpha}\nabla(\delta_{\hat{\rho_\alpha}}\hat{V})-\hat{\rho}_{\beta}\nabla(\delta_{\hat{\rho_\beta}}\hat{V}) \enskip , \quad
\end{equation*}
which still preserves the momentum and implies the following coarse-grained equation:
\begin{equation}
\label{AppAbin}
\partial_t (\rho\mathbf{v}) + \nabla\cdot(\rho\mathbf{v}\otimes\mathbf{v}) = -\rho_{\alpha}\nabla(\delta_{\rho_\alpha}F)-\rho_{\beta}\nabla(\delta_{\rho_\beta}F) \enskip .
\end{equation}
It is straightforward to show, that if the free energy is the functional of the densities and their spatial gradients, there exists $\mathbb{R}$ such that $\nabla\cdot\mathbb{R}$ equals the righ-hand side of Eq. (\ref{AppAbin}) (see Appendix B). Finally, we interpret the non-equilibrium solid-liquid system as a reactive binary system, and we adopt:
\begin{equation*}
\partial_t (\rho\mathbf{v}) + \nabla\cdot(\rho\mathbf{v}\otimes\mathbf{v}) := -\rho_{s}\nabla(\delta_{\rho_s}F)-\rho_{l}\nabla(\delta_{\rho_l}F) \enskip , \quad
\end{equation*}
which also becomes $0$ in equilibrium [described by $\delta_{\rho_s} F = \delta_{\rho_l} F = \mu$ (constant)], but preserves the full momentum in non-equilibrium.

\subsection*{B. Constitutive relations and variable transformation}

\subsubsection*{The Reversible stress tensor}

The general form of an Navier-Stokes equation that conserves the full momentum of the system reads:
\begin{equation*}
\partial_t (\rho\mathbf{v})+\nabla\cdot(\rho\,\mathbf{v}\otimes\mathbf{v}) = \nabla\cdot\mathbb{R} \enskip ,
\end{equation*}
where $\mathbb{R}$ is the reversible stress tensor. If the spatio-temporal state of the system is described by only the density [$\rho(\mathbf{r},t)$] and the velocity [$\mathbf{v}(\mathbf{r},t)$] fields, the following relation applies:
\begin{equation*}
\nabla\cdot\mathbb{R} = -\rho \nabla(\delta_\rho F[\rho]) \enskip ,
\end{equation*}
which is often called generalized Gibbs-Duhem relation. Assuming that the free energy density $f$ depends only on the density and its gradient, $\mathbb{R}$ can be found in 1 spatial dimension by integrating the above equation:
\begin{equation*}
\begin{split}
R = & - \int \rho (\delta_\rho F)' = - \rho (\delta_\rho F) + \int \rho'(\delta_\rho F) = \\
& - \rho (\delta_\rho F) + \int \rho'[\partial_\rho f - (\partial_{\rho'}f)'] =\\
& f - \rho (\delta_\rho F) - \rho' (\partial_{\rho'}f) \enskip ,
\end{split} 
\end{equation*}
where $()'$ stands for $d/dx$ and we used that $f'=(\partial_\rho f)\rho'+(\partial_{\rho'} f)\rho''$. The multi-dimensional generalization of the above result reads:
\begin{equation*}
\mathbb{R} = [ f - \rho (\delta_\rho F)]\,\mathbb{I}-\nabla\rho \otimes (\partial_{\nabla\rho}f)  \enskip .
\end{equation*}
The calculation can be repeated for $F[\rho_s,\rho_l]$ with
\begin{equation}
\label{AppBGD1}\nabla\cdot\mathbb{R} = -\rho_s \nabla(\delta_{\rho_s} F[\rho_s,\rho_l]) -\rho_l \nabla(\delta_{\rho_l} F[\rho_s,\rho_l]) \enskip ,
\end{equation}
yielding:
\begin{equation}
\label{AppBR1}
\begin{split}
\mathbb{R} = & [f-\rho_s (\delta_{\rho_s}F)-\rho_l (\delta_{\rho_l}F)]\,\mathbb{I} \\
& - \nabla\rho_s \otimes (\partial_{\nabla\rho_s}f)- \nabla\rho_l \otimes (\partial_{\nabla\rho_l}f) \enskip ,
\end{split}
\end{equation}
where we used that $f'=(\partial_{\rho_s} f)\rho_s'+(\partial_{\rho_l} f)\rho_l'+(\partial_{\rho_s'} f)\rho_s''+(\partial_{\rho_l'} f)\rho_l''$. The results show that Eq. (\ref{AppBGD1}) preserves the full momentum for $F[\rho_s,\rho_l]$.\\

One can also try to combine $\nabla\cdot\mathbb{R}=-\rho\nabla(\delta_\rho F)$ and $F[\rho,\phi]$. We assume that such a $g$ exists, for which $R:=f-\rho(\delta_\rho F)-\rho' \partial_{\rho'}f-g$ result in $R'=-\rho(\delta_\rho F)'$. The differentiation of $R$ yields:
\begin{equation*}
R' = -\rho(\delta_\rho F)' + [ \phi'(\partial_\phi f) + \phi''(\partial_{\phi'} f) - g' ] \enskip ,
\end{equation*}
thus indicating $\phi'(\partial_\phi f) + \phi''(\partial_{\phi'}f) = g'$, and therefore
\begin{equation*}
g = \int [\phi'(\partial_\phi f) + \phi''(\partial_{\phi'}f)] \enskip ,
\end{equation*}
or, equivalently:
\begin{equation*}
g = f - \int [\rho'(\partial_\rho f) + \rho''(\partial_{\rho'}f)] \enskip ,
\end{equation*}
where we used $f'=(\partial_\rho f)\rho'+(\partial_{\rho'} f)\rho''+(\partial_\phi f)\phi'+(\partial_{\phi'} f)\phi''$. In simple gradient theories the general form of the free energy density reads: $f=f_0(\phi,\rho)+(a/2)(\phi')^2+(b/2)(\rho')^2$, and therefore $\int \phi''(\partial_{\phi'}f) = a (\phi')^2$ and $\int \rho''(\partial_{\rho'}f) = b (\rho')^2$. The problem is, that $\int \phi'(\partial_\phi f_0)$ and $\int \rho'(\partial_\rho f_0)$ are integrable (by substitution) only for $f_0(\phi,\rho)=p(\phi)+q(\rho)$, i.e., when the phase and the density are decoupled. Consequently, if $f_0(\phi,\rho)$ contains any coupling between $\phi$ and $\rho$, no $\mathbb{R}$ satisfying $\nabla\cdot\mathbb{R}=-\rho\nabla(\delta_\rho F[\rho,\phi])$ exists. 

\subsubsection*{Mass fraction formalism}

A crucial point of the model development is a rigorous definition of the phase-field. Without an exact definition, the dynamical equation for the phase-field might be incidental, which we would like to avoid. The dynamical equations of the solid-liquid system read:
\begin{eqnarray}
\nonumber\partial_t \rho_s + \nabla\cdot(\rho_s\mathbf{v}) &=& -M \left( \delta_{\rho_s}F - \delta_{\rho_l}F \right) \enskip ;\\
\label{AppBsum}\partial_t \rho + \nabla\cdot(\rho\mathbf{v}) &=& 0 \enskip ;\\
\nonumber\partial_t (\rho\mathbf{v}) + \nabla\cdot(\rho\mathbf{v}\otimes\mathbf{v}) &=& -\rho_s \nabla(\partial_{\rho_s}F)-\rho_l \nabla(\partial_{\rho_l}F) \enskip ,
\end{eqnarray}
where $\rho=\rho_s+\rho_l$, and therefore one of the 3 variables is reducible. The first step is to re-write the above equations in terms of the mass fractions and the total density. Our variables are $c_s:=\rho_s/\rho$, $c_l:=\rho_l/\rho$, and $\rho=\rho_s+\rho_l$. Our task is to express the right-hand sides in Eq. (\ref{AppBsum}) in terms of $c_s$, $c_l$ and $\rho$. This can be done by applying the chain rule for the functional derivatives \cite{PhysRevB.92.184105}. Introducing $\tilde{F}[\rho,c_s,c_l]:=F[c_s\rho,c_l\rho]$ results in:
\begin{equation*}
\begin{split}
\frac{\delta F}{\delta\rho_s}= &\frac{\delta \tilde{F}}{\delta \rho}\frac{\partial \rho}{\partial\rho_s}+\frac{\delta \tilde{F}}{\delta c_s}\frac{\partial c_s}{\partial\rho_s}+\frac{\delta \tilde{F}}{\delta c_l}\frac{\partial c_l}{\partial \rho_s} = \\
& \frac{\delta \tilde{F}}{\delta \rho}+\frac{1}{\rho}\left(\frac{\delta \tilde{F}}{\delta c_s}-\Sigma \right) \enskip ,
\end{split}
\end{equation*}
where $\Sigma = c_s\frac{\delta \tilde{F}}{\delta c_s}+c_l\frac{\delta \tilde{F}}{\delta c_l}$, and we used that $\frac{\partial \rho}{\partial \rho_s}=1$, $\frac{\partial c_s}{\partial \rho_s}=\frac{1-c_s}{\rho}$, and $\frac{\partial c_l}{\partial \rho_s}=-\frac{c_l}{\rho}$. Similarly,
\begin{equation*}
\begin{split}
\frac{\delta F}{\delta\rho_l}= \frac{\delta \tilde{F}}{\delta \rho}+\frac{1}{\rho}\left(\frac{\delta \tilde{F}}{\delta c_l}-\Sigma \right) \enskip .
\end{split}
\end{equation*}
Using these in Eq. (\ref{AppBsum}) results in:
\begin{eqnarray*}
\frac{\delta F}{\delta\rho_s}-\frac{\delta F}{\delta\rho_l} &=& \frac{1}{\rho}\left(\frac{\delta \tilde{F}}{\delta c_s}-\frac{\delta \tilde{F}}{\delta c_l} \right) \enskip ;\\
-\rho_s\nabla\frac{\delta F}{\delta\rho_s}-\rho_l\nabla\frac{\delta F}{\delta\rho_l}&=&-\rho\nabla\frac{\delta \tilde{F}}{\delta \rho}+\frac{\delta \tilde{F}}{\delta c_s}\nabla c_s+\frac{\delta \tilde{F}}{\delta c_l}\nabla c_l \enskip ,
\end{eqnarray*}
where the second equation emerges from:
\begin{equation*}
\begin{split}
& -\rho_s\nabla\frac{\delta F}{\delta\rho_s} -\rho_l\nabla\frac{\delta F}{\delta\rho_l} = \\
& -c_s \rho \nabla\left[ \frac{\delta \tilde{F}}{\delta\rho} + \frac{1}{\rho} \left( \frac{\delta \tilde{F}}{\delta c_s} - \Sigma \right) \right] \\
& -c_l \rho \nabla\left[ \frac{\delta \tilde{F}}{\delta\rho} + \frac{1}{\rho} \left( \frac{\delta \tilde{F}}{\delta c_l} - \Sigma \right) \right] =\\
 & - \rho \nabla \frac{\delta \tilde{F}}{\delta \rho} -c_s \nabla\left( \frac{\delta \tilde{F}}{\delta c_s} - \Sigma \right) -c_l \nabla\left( \frac{\delta \tilde{F}}{\delta c_l} - \Sigma \right) = \\
& - \rho \nabla \frac{\delta \tilde{F}}{\delta \rho} +  \frac{\delta \tilde{F}}{\delta c_s} \nabla c_s +  \frac{\delta \tilde{F}}{\delta c_l}\nabla c_l \enskip .
\end{split}
\end{equation*}
In the final step we introduce $\tilde{\tilde{F}}[\rho,\phi]:=F[\rho,\phi,1-\phi]$ to reduce the liquid mass fraction in $\tilde{F}[\rho,c_s,c_l]$, yielding:
\begin{eqnarray*}
\left(\frac{\delta \tilde{F}}{\delta c_s}-\frac{\delta \tilde{F}}{\delta c_l} \right) &=& \frac{\delta \tilde{\tilde{F}}}{\delta\phi} ;\\
\frac{\delta \tilde{F}}{\delta c_s}\nabla c_s+\frac{\delta \tilde{F}}{\delta c_l}\nabla c_l &=&\frac{\delta \tilde{\tilde{F}}}{\delta \phi}\nabla \phi \enskip .
\end{eqnarray*}
Using these together with the continuity equation in Eq. (\ref{AppBsum}) results in:
\begin{eqnarray*}
\rho \dot{c}_s &=& -(M/\rho) (\delta_\phi F]) \enskip ;\\
\dot{\rho} &=& - \rho (\nabla\cdot\mathbf{v}) \enskip ;\\
\rho \dot{\mathbf{v}} &=& -\rho\nabla(\delta_\rho F+(\delta_\phi F)\nabla\phi \enskip ,
\end{eqnarray*}
where $\dot{()}=\partial_t+\mathbf{v}\cdot\nabla()$ stands for the material derivative, and we omitted the double tilde notation. Furthermore, using the variable transformation in Eq. (\ref{AppBR1}) results in:
\begin{equation*}
\mathbb{R} = [ f - \rho (\delta_\rho F)]\,\mathbb{I}-[\nabla\rho \otimes (\partial_{\nabla\rho}f)+\nabla\phi \otimes (\partial_{\nabla\phi}f)]  \enskip ,
\end{equation*}
which indeed results in
\begin{equation}
\label{AppBGDfinal}
\nabla \cdot \mathbb{R} =  -\rho\nabla(\delta_\rho F)+(\delta_\phi F)\nabla\phi \enskip .
\end{equation}
Finally, the Euler Lagrange equations transform as follows. Using the variable transformation in the original Euler-Lagrange equations results in:
\begin{eqnarray}
\label{AppBEL1}\frac{\delta F}{\delta \rho_s} = \frac{\delta F}{\delta \rho}+\frac{1}{\rho}\left( \frac{\delta F}{\delta c_s} - \Sigma \right) = \mu \enskip ; \\
\label{AppBEL2}\frac{\delta F}{\delta \rho_l} = \frac{\delta F}{\delta \rho}+\frac{1}{\rho}\left( \frac{\delta F}{\delta c_l} - \Sigma \right) = \mu \enskip .
\end{eqnarray}
Multiplying the first equation by $c_s$ and the second by $c_l$, then adding the results yields:
\begin{equation}
\label{AppBEL1f}
\delta_\rho F = \mu .
\end{equation}
Using this in the Equations (\ref{AppBEL1}) and (\ref{AppBEL2}) gives:
\begin{equation*}
\left( \begin{matrix} 1-c_s & - c_l \\ -c_s & 1-c_l \end{matrix} \right) \left( \begin{matrix} \delta_{c_s} F \\ \delta_{c_l} F \end{matrix} \right) = \left( \begin{matrix} 0 \\ 0\end{matrix} \right) \enskip .
\end{equation*}
Given that $c_l+c_s=1$, the coefficient matrix has two eigenvalues, namely, $0$ and $1$. The eigenvector for $0$ is $(1,1)$, which indicates, that $\delta_{c_s}F=\delta_{c_l}F$ satisfies the above equation independently from $c_s$ and $c_l$. The Euler-Lagrange equation for the phase then reads:
\begin{equation}
\label{AppBEL2f}
\delta_{c_s}F = \delta_{c_l}F \quad \Rightarrow \quad \delta_\phi F = 0 \enskip . 
\end{equation}
Note that Equations (\ref{AppBEL1f}) and (\ref{AppBEL2f}) indicate that Eq. (\ref{AppBGDfinal}) becomes $0$ in equilibrium.

\subsection*{C. The phase mobility}

The only fitting parameter of our model is the phase-field mobility, $\Gamma$, which determines the characteristic time scale of phase relaxation. To estimate $\Gamma$, first we solve the reduced phase-field equation
\begin{equation}
\label{AppCphase}
\bar{\rho}\,\partial_t \phi = -\Gamma_0 (\delta_\phi F_0) 
\end{equation}
for a propagating planar solid-liquid interface, where $\delta_\phi F_0=\frac{\sigma}{\delta}\frac{dg}{d\phi}+\Lambda\frac{dp}{d\phi}-(3\,\sigma\,\delta)\partial_x^2\phi$, and $\Gamma_0$ constant. The 1-dimensional Euler-Lagrange equation (referring to $\Lambda=0$) prescribes:
\begin{equation*}
\frac{\delta F_0}{\delta \phi} = \frac{\sigma}{\delta} \left[ \left.\frac{dg(\phi)}{d\phi}\right|_{\phi=\phi_0(x)}-3\,\delta^2\,\phi_0''(x) \right] = 0 \enskip ,
\end{equation*} 
which provides the well-known interface solution $\phi_0(x)=[1-\tanh(x/\delta)]/2$. Substituting $\phi(x,t):=\phi_0(x-V\,t)$ into Eq. (\ref{AppCphase}) results in:
\begin{equation*}
(3\,\Gamma_0\,\delta\,\Lambda+V\,\bar{\rho})\text{sech}(x-V\,t)=0 \enskip ,
\end{equation*}
which indicates the following front speed:
\begin{equation}
\label{AppCVPFT}
V_{PFT} = -\frac{3\,\Gamma_0\,\delta}{\bar{\rho}}\Lambda \enskip .
\end{equation}
On the other hand, Wilson and Frenkel gave the following expression for the front speed:
\begin{equation*}
V_{WF} = -\frac{D a_0}{\ell^2}\left[ 1-e^{-\Delta G/(R\,T)} \right] \enskip ,
\end{equation*}
where $D$, $a_0$ and $\ell$ are the self-diffusion coefficient, the inter-atomic spacing, and the diffusional mean free path in the liquid, respectively, while $\Delta G = \Lambda\,v_M$ (where $v_M$ is the molar volume), and $R$ the gas constant. For $|\Lambda|\ll 1$, the Wilson-Frenkel gives $V_{WF} \approx -\frac{D\,a_0}{\ell^2}\frac{v_M}{R\,T} \Lambda$. Equating this and the right-hand side of Eq. (\ref{AppCVPFT}) results in:
\begin{equation*}
\Gamma_0 = \frac{\bar{\rho}}{3}\frac{D\,a_0}{\delta\,\ell^2}\frac{v_M}{R\,T} \enskip .
\end{equation*}
The dimensionless phase-field mobility then reads:
\begin{equation*}
\kappa_0 = \Gamma_0 \sqrt{\frac{\sigma\,\delta}{\bar{\rho}}} = \frac{D a_0}{3\,\ell^2} \frac{v_M}{R\,T}\sqrt{\frac{\sigma\,\bar{\rho}}{\delta}}  \enskip ,
\end{equation*}
while the dimensionless front speed is:
\begin{equation*}
\tilde{V} = V \frac{\tau}{\delta} = -\frac{3\,\Gamma_0 \,\delta\,\Lambda}{\bar{\rho}}\frac{1}{\delta}\sqrt{\frac{\bar{\rho}\,\delta^3}{\sigma}} = -3\,\kappa_0\,\lambda \enskip ,
\end{equation*}
where we used $\lambda=(\delta/\sigma)\Lambda$ and $\kappa_0 = \Gamma_0 \sqrt{\sigma\,\delta/\bar{\rho}}$.

\subsection*{D. Quasi-incompressible hydrodynamics}

To derive quasi-incompressible hydrodynamic equations, we start from the compressible dynamical equations
\begin{eqnarray*}
\rho \, \dot{\phi} &=& -\kappa(\delta_\phi F) \\
\dot{\rho} &=& -\rho(\nabla\cdot\mathbf{v}) \\
\rho \, \dot{\mathbf{v}} &=& -\rho\nabla(\delta_\rho F)+(\delta_\phi F)\nabla\phi \enskip ,
\end{eqnarray*}
supplemented with the local condition
\begin{equation}
\label{AppDcond}
\rho(\mathbf{r},t)=h[\phi(\mathbf{r},t)] \enskip ,
\end{equation}
where $\rho_0(x)=h[\phi_0(x)]$ (equilibrium planar interface). Imposing Eq. (\ref{AppDcond}) on the dynamical equations means that the density is fixed as a function of the phase also in non-equilibrium, and is therefore not a dependent variable any more. Consequently, the dynamics is over-determined, which, however, can be resolved by utilizing the Lagrange multiplier method. In the presence of the local condition the Euler-Lagrange equations read:
\begin{eqnarray*}
\frac{\delta F}{\delta \phi} - \theta_0(\mathbf{r})\frac{\partial w(\phi,\rho)}{\partial \phi} &=& 0 \\
\frac{\delta F}{\delta \rho} - \theta_0(\mathbf{r})\frac{\partial w(\phi,\rho)}{\partial \rho} &=& \mu_0 \enskip ,
\end{eqnarray*}
where $w(\phi,\rho):=\rho-h[\phi]=0$, and the Lagrange multiplier $\theta_0(\mathbf{r})$ is the function of $\mathbf{r}$ (the condition is local). Plugging the equilibrium planar interface solution into the above equations yields:
\begin{equation*}
\theta_0(\mathbf{r})\left.\frac{dh(\phi)}{d\phi}\right|_{\phi=\phi_0(x)} = 0 \quad \text{and} \quad -\theta_0(\mathbf{r})=0 \enskip ,
\end{equation*}
where we used that $\delta_\phi F = 0$ and $\delta_\rho F = \mu$ at $\phi_0(x)$ and $\rho_0(x)$. Consequently, the Lagrange multiplier is identically $0$ in equilibrium, which accords with the fact that Eq. (\ref{AppDcond}) represent no extra condition in equilibrium. Imposing the local condition on the dynamical equations, the phase-field and continuity equations transform as follows:
\begin{eqnarray*}
h(\phi) \, \dot{\phi} &=& -\kappa \left( \delta_\phi^{(c)} F \right) \enskip ; \\
\nabla\cdot\mathbf{v} &=& - \frac{\dot{\rho}}{\rho} = - \frac{1}{h(\phi)}\frac{dh(\phi)}{d\phi}\dot{\phi} = \frac{\kappa}{h^2(\phi)}\frac{dh(\phi)}{d\phi}\left(\delta_\phi^{(c)} F \right) \enskip ,
\end{eqnarray*}
where $\left(\delta_\phi^{(c)} F \right) =\delta_\phi F|_{\rho=h(\phi)}$ is the "conditional" functional derivative, which only depends on $\phi$. Note that $\theta(\mathbf{r},t)$ is not included in these equations. to resolve the continuity equation, which defines the divergence of the velocity field as a function of the phase field, The Lagrange multiplier must be included in the Navier-Stokes equation. In this, we use the substitutions $\delta_\rho F \to \left(\delta_\rho^{(c)} F \right) - \theta(\mathbf{r},t)(\partial_\rho w)$ and $\delta_\phi F \to \left( \delta_\phi^{(c)} F \right) - \theta(\mathbf{r},t)(\partial_\phi w)$, yielding:
\begin{equation*}
\begin{split}
&-\rho\nabla\left[ \left(\delta_\rho^{(c)} F \right) - \theta(\mathbf{r},t)(\partial_\rho w) \right] \\
&+ \left[ \left( \delta_\phi^{(c)} F \right) - \theta(\mathbf{r},t)(\partial_\rho w) \right] \nabla\phi = \\
& -\rho\nabla \left(\delta_\rho^{(c)} F \right)+ \left( \delta_\phi^{(c)} F \right)\nabla\phi + \nabla[\rho\,\theta\,(\partial_\rho w)] \enskip ,
\end{split}
\end{equation*}
where we used that $\nabla w = (\partial_\phi w)\nabla\phi+(\partial_\rho w)\nabla\rho=0$. Using $(\partial_\rho w)=1$, $\left(\delta_\rho^{(c)} F \right)=\delta_\rho F|_{\rho=h(\phi)}=\mu_0$ (constant), and introducing $\delta p:=-\rho\,\theta$ yield:
\begin{equation*}
h(\phi) \, \dot{\mathbf{v}} = \left(\delta_\phi^{(c)} F \right) \nabla\phi -\nabla\delta p \enskip ,
\end{equation*}
which is a usual incompressible Navier-Stokes equation, where $\delta p(\mathbf{r},t)$ is present to guarantee continuity, and can be resolved in a numerical simulation by applying Chorin's projection method, for instance.

\subsection*{E. Coexistence densities and equilibrium planar interfaces}

To find the densities of two coexisting phases, we need to start from the 1-dimensional Euler-Lagrange equations, which read:
\begin{equation*}
\delta_\phi F = 0 \quad \text{and} \quad \delta_\rho F = \mu_0 \enskip ,
\end{equation*}
where $\mu_0$ is the (unknown) equilibrium chemical potential. The boundary conditions read:
\begin{equation*}
\lim_{x \to \pm\infty}\phi_0(x) = \phi_{1,2} \quad \text{and} \quad \lim_{x \to \pm\infty}\rho_0(x) = \rho_{1,2} \enskip ,
\end{equation*}
where $\phi_{1,2}$ are two different homogeneous solutions of the equation $\delta_\phi F=0$ for arbitrary constant density. The Euler-Lagrange equations can be expanded, yielding:
\begin{eqnarray}
\nonumber 3 \phi_0''(x) = \left[ \frac{dg(\phi)}{d\phi} + B \frac{\partial f_\rho(\phi,\rho)}{\partial \phi} \right]_{\phi_0(x),\rho_0(x)} \enskip ;\\
\label{AppErho}\mu_0 = \left[ B \frac{\partial f_\rho(\phi,\rho)}{\partial \rho} \right]_{\phi_0(x),\rho_0(x)} \enskip .
\end{eqnarray}
Multiplying the first equation by $\phi_0'(x)$ and the second by $\rho_0'(x)$, then adding the results yields:
\begin{equation}
\label{AppEDW}
\frac{d}{dx}\left[\frac{3}{2}(\phi_0')^2 + \mu_0\rho_0 \right] = \frac{d}{dx}\left[ g(\phi_0)+B f_\rho(\phi_0,\rho_0) \right] \enskip .
\end{equation}
Integrating the above equation from $x \to -\infty$ to $x \to +\infty$ results in:
\begin{equation}
\label{AppEdw}
\mu_0(\rho_2-\rho_1)= f_2(\rho_2) - f_1(\rho_1) \enskip , 
\end{equation}
where $f_i(\rho)= g(\phi_i)+B f_\rho(\phi_i,\rho)$. Taking the limits $x \to \pm \infty$ for Eq. (\ref{AppErho}) yields:
\begin{equation}
\label{AppEchem}
\mu_0= \mu_2(\rho_2) = \mu_1(\rho_1) \enskip , 
\end{equation}
where $\mu_i(\rho)=B [\partial_\rho f_\rho(\phi,\rho)]_{\phi=\phi_i}$. Eliminating $\mu_0$ from Equations (\ref{AppEchem}) and (\ref{AppEdw}) results in:
\begin{eqnarray}
\label{AppEct1}\mu_2(\rho_2) &=& \mu_1(\rho_1) \enskip ;\\
\label{AppEct2}f_1(\rho_1)-\rho_1\mu_1(\rho_1) &=& f_2(\rho_2) -\rho_2\mu_2(\rho_2) \enskip ,
\end{eqnarray}
which are two equations for two unknowns, $\rho_1$ and $\rho_2$. Note that as long as $\phi_1$ and $\phi_2$ don't depend on $\rho$, $\mu_i(\rho) \equiv df_i(\rho)/d\rho$, and therefore Equations (\ref{AppEct1}) and (\ref{AppEct2}) give the common tangent construction. This applies to $\phi=0$ and $\phi=1$, but the other homogeneous solution(s) of $\delta_\phi F = 0$ usually depend(s) on the density, in which case $\mu_i(\rho)\neq df_i(\rho)/d\rho$, and therefore the common tangent representation is not applicable. To find the planar interface solution in equilibrium, first we express the density as a function of the phase field from the Euler-Lagrange equation for the density, where the equilibrium chemical potential reads:
\begin{equation*}
B\partial_\rho f_\rho = \mu_0 \quad \Rightarrow \quad \rho = h(\phi) \enskip .
\end{equation*}
Next, we substitute $\rho$ by $h(\phi)$ in the Euler-Lagrange equation $\delta_\phi F=0$, which yields:
\begin{equation*}
\phi_0''(x) = s[\phi_0(x)] \enskip ,
\end{equation*}
where $s(\phi)=\frac{1}{3}\left[ \frac{dg(\phi)}{d\phi} + B \left. \frac{\partial f_\rho(\phi,\rho)}{\partial \phi} \right|_{\rho=h(\phi)} \right]$. Multiplying the above equation by $\phi_0'(x)$, then integrating it from $x \to -\infty$ to $x = z$ results in:
\begin{equation}
\label{AppEprof}
\frac{1}{2}[\phi_0'(z)]^2 = S[\phi_0(z)] \enskip ,
\end{equation}
where $S(\phi) = \int_{\phi_0}^\phi d\psi s(\psi)$. Separating the variables and integrating again yields:
\begin{equation*}
z = \int \frac{d\phi}{\sqrt{2\,S(\phi)}} \enskip ,
\end{equation*}
which is an implicit equation for $\phi_0(z)$. If $S(\phi)$ exist, the integral on the right-hand side of the above equation can be evaluated analytically, \text{and} the equation can be inverted, we have an explicit analytical expression for the phase-field profile, while the density profile is simply given by $\rho_0(z)=h[\phi_0(z)]$. The interfacial tension (the excess energy of the equilibrium planar interface) can be calculated \textit{without} knowing the interface. The definition of the interfacial tension reads:
\begin{equation*}
f_\sigma = \int_{-\infty}^{+\infty} dz \left\{ f-f_0-\mu_0(\rho-\rho_0) \right\} \enskip ,
\end{equation*}
where $f=\frac{3}{2}[\phi_0'(z)]^2+\Delta f[\phi_0(z),\rho_0(z)]$, and therefore $\Delta f(\phi,\rho)=g(\phi)+B f_\rho(\phi,\rho)$, $\mu_0$ is the equilibrium chemical potential, $f_0 = \lim_{z \to -\infty} f$, and $\rho_0 = \lim_{z \to -\infty} \rho_0(z)$. Note that taking the integrand of the above integral in the $z \to +\infty$ limit recovers Eq. (\ref{AppEct2}). Integrating Eq. (\ref{AppEDW}) from $x \to -\infty$ to $x=z$ yields:
\begin{equation*}
\frac{3}{2}[\phi_0'(z)]^2 = \Delta f[\phi_0(z),\rho_0(z)] - f_0 - \mu_0[\rho_0(z)-\rho_0] \enskip ,
\end{equation*}
thus indicating
\begin{equation*}
f_\sigma = 3 \int_{-\infty}^{+\infty} dz [\phi_0'(z)]^2 = 3 \int_{\phi_0}^{\phi_1} \sqrt{2\,S(\phi)} \enskip , 
\end{equation*}
where we used Equation (\ref{AppEprof}).
 
\subsection*{Appendix F: Front velocity in the Oxtoby-Harrowell model}

In Reference \cite{doi:10.1063/1.462864}  the authors used the following dynamical equations:
\begin{eqnarray*}
n\,\partial_t \phi &=& - [\gamma/(k_B T)]\delta_\phi F[\phi,n] \enskip ; \\
\partial_t \rho + \nabla\cdot(\rho\mathbf{v}) &=& 0 \enskip ; \\
n\,(\partial_t \mathbf{v}+\mathbf{v}\cdot\nabla\mathbf{v})&=&-n \,\delta_n F[\phi,n] + \nu\,\nabla\cdot\mathbb{D} \enskip ,
\end{eqnarray*}
where $n$ was the number density, $\phi$ the solid order parameter (phase field), $\gamma$ a constant, $\nu$ the viscosity (in atomic mass units), and $\mathbb{D}$ the viscous stress. The free energy functional was defined as:
\begin{equation*}
\frac{F[\phi,n]}{k_B T n_s} = \int dV \left\{ \omega(\phi,n) + \frac{K_m^2}{2}(\nabla \phi)^2 + \frac{K_n}{2}(\nabla n)^2 \right\} \enskip ,
\end{equation*} 
where $n_s$ is the solid coexistence density, and
\begin{equation*}
\begin{split}
\omega(\phi,n) = \min & \left\{ \lambda_1\frac{\phi^2}{2} + \frac{\lambda_2}{2}\left(\frac{n-n_l}{n_l}\right)^2,\right. \\
& \left. \lambda_3\frac{(\phi-\phi_0)^2}{2} + \frac{\lambda_4}{2}\left(\frac{n-n_s}{n_s}\right)^2 + \Delta \right\} \enskip ,
\end{split}
\end{equation*}
where $n_l$ is the liquid coexistence density, $\phi_0$ the value of the phase-field in the bulk solid, and $\Delta$ the dimensionless thermodynamic driving force. In the inviscid limit the dimensional front velocity reads:
\begin{equation}
\label{AppF:OHV}
v = -\gamma\,\Delta\,K_m\,\frac{2}{\phi_0^2}\frac{\sqrt{\lambda_1}+\sqrt{\lambda_3}}{\sqrt{\lambda_1\lambda_3}} \enskip .
\end{equation}
Setting $\phi_0=1$ and $\lambda_0:=\lambda_1=\lambda_3$, denoting the atomic mass by $M$, and comparing the above equations to our model results in the following parameters:
\begin{eqnarray*}
\lambda_0 &=& \frac{12}{k_B T n_s}\frac{\sigma}{\delta}\\
\Delta &=& \Lambda/(n_s k_B T) \\
K_m &=& \sqrt{3\,\sigma\,\delta/(n_s k_B T)}\\
\gamma &=& k_B T \,\Gamma_0 / M  
\end{eqnarray*}
Using these in Eq. (\ref{AppF:OHV}) yields $v = -2\,\Gamma_0\,\Lambda\,\delta/\rho_s$, which accords with Eq. (\ref{central}) up to a constant factor. The difference emerges from the fact that the form of $\omega(\phi,n)$ is different for the two models. 

\bibliography{./fluids}

\begin{thebibliography}{33}%
\makeatletter
\providecommand \@ifxundefined [1]{%
 \@ifx{#1\undefined}
}%
\providecommand \@ifnum [1]{%
 \ifnum #1\expandafter \@firstoftwo
 \else \expandafter \@secondoftwo
 \fi
}%
\providecommand \@ifx [1]{%
 \ifx #1\expandafter \@firstoftwo
 \else \expandafter \@secondoftwo
 \fi
}%
\providecommand \natexlab [1]{#1}%
\providecommand \enquote  [1]{``#1''}%
\providecommand \bibnamefont  [1]{#1}%
\providecommand \bibfnamefont [1]{#1}%
\providecommand \citenamefont [1]{#1}%
\providecommand \href@noop [0]{\@secondoftwo}%
\providecommand \href [0]{\begingroup \@sanitize@url \@href}%
\providecommand \@href[1]{\@@startlink{#1}\@@href}%
\providecommand \@@href[1]{\endgroup#1\@@endlink}%
\providecommand \@sanitize@url [0]{\catcode `\\12\catcode `\$12\catcode
  `\&12\catcode `\#12\catcode `\^12\catcode `\_12\catcode `\%12\relax}%
\providecommand \@@startlink[1]{}%
\providecommand \@@endlink[0]{}%
\providecommand \url  [0]{\begingroup\@sanitize@url \@url }%
\providecommand \@url [1]{\endgroup\@href {#1}{\urlprefix }}%
\providecommand \urlprefix  [0]{URL }%
\providecommand \Eprint [0]{\href }%
\providecommand \doibase [0]{http://dx.doi.org/}%
\providecommand \selectlanguage [0]{\@gobble}%
\providecommand \bibinfo  [0]{\@secondoftwo}%
\providecommand \bibfield  [0]{\@secondoftwo}%
\providecommand \translation [1]{[#1]}%
\providecommand \BibitemOpen [0]{}%
\providecommand \bibitemStop [0]{}%
\providecommand \bibitemNoStop [0]{.\EOS\space}%
\providecommand \EOS [0]{\spacefactor3000\relax}%
\providecommand \BibitemShut  [1]{\csname bibitem#1\endcsname}%
\let\auto@bib@innerbib\@empty
%</preamble>
\bibitem [{\citenamefont {Glicksman}\ \emph {et~al.}(1994)\citenamefont
  {Glicksman}, \citenamefont {Koss},\ and\ \citenamefont
  {Winsa}}]{PhysRevLett.73.573}%
  \BibitemOpen
  \bibfield  {author} {\bibinfo {author} {\bibfnamefont {M.~E.}\ \bibnamefont
  {Glicksman}}, \bibinfo {author} {\bibfnamefont {M.~B.}\ \bibnamefont {Koss}},
  \ and\ \bibinfo {author} {\bibfnamefont {E.~A.}\ \bibnamefont {Winsa}},\
  }\href {\doibase 10.1103/PhysRevLett.73.573} {\bibfield  {journal} {\bibinfo
  {journal} {Phys. Rev. Lett.}\ }\textbf {\bibinfo {volume} {73}},\ \bibinfo
  {pages} {573} (\bibinfo {year} {1994})}\BibitemShut {NoStop}%
\bibitem [{\citenamefont {Pines}\ \emph {et~al.}(1996)\citenamefont {Pines},
  \citenamefont {Chait},\ and\ \citenamefont {Zlatkowski}}]{PINES1996798}%
  \BibitemOpen
  \bibfield  {author} {\bibinfo {author} {\bibfnamefont {V.}~\bibnamefont
  {Pines}}, \bibinfo {author} {\bibfnamefont {A.}~\bibnamefont {Chait}}, \ and\
  \bibinfo {author} {\bibfnamefont {M.}~\bibnamefont {Zlatkowski}},\ }\href
  {\doibase https://doi.org/10.1016/S0022-0248(96)00667-7} {\bibfield
  {journal} {\bibinfo  {journal} {Journal of Crystal Growth}\ }\textbf
  {\bibinfo {volume} {169}},\ \bibinfo {pages} {798 } (\bibinfo {year}
  {1996})}\BibitemShut {NoStop}%
\bibitem [{\citenamefont {Ivantsov}(1947)}]{Ivantsov}%
  \BibitemOpen
  \bibfield  {author} {\bibinfo {author} {\bibfnamefont {G.}~\bibnamefont
  {Ivantsov}},\ }\href@noop {} {\bibfield  {journal} {\bibinfo  {journal}
  {DokI. Akad. Nauk SSSR}\ }\textbf {\bibinfo {volume} {58}},\ \bibinfo {pages}
  {567} (\bibinfo {year} {1947})}\BibitemShut {NoStop}%
\bibitem [{\citenamefont {McFadden}\ and\ \citenamefont
  {Coriell}(1986)}]{MCFADDEN1986507}%
  \BibitemOpen
  \bibfield  {author} {\bibinfo {author} {\bibfnamefont {G.}~\bibnamefont
  {McFadden}}\ and\ \bibinfo {author} {\bibfnamefont {S.}~\bibnamefont
  {Coriell}},\ }\href {\doibase https://doi.org/10.1016/0022-0248(86)90195-8}
  {\bibfield  {journal} {\bibinfo  {journal} {Journal of Crystal Growth}\
  }\textbf {\bibinfo {volume} {74}},\ \bibinfo {pages} {507 } (\bibinfo {year}
  {1986})}\BibitemShut {NoStop}%
\bibitem [{\citenamefont {Horvay}(1965)}]{HORVAY1965195}%
  \BibitemOpen
  \bibfield  {author} {\bibinfo {author} {\bibfnamefont {G.}~\bibnamefont
  {Horvay}},\ }\href {\doibase https://doi.org/10.1016/0017-9310(65)90110-9}
  {\bibfield  {journal} {\bibinfo  {journal} {International Journal of Heat and
  Mass Transfer}\ }\textbf {\bibinfo {volume} {8}},\ \bibinfo {pages} {195 }
  (\bibinfo {year} {1965})}\BibitemShut {NoStop}%
\bibitem [{\citenamefont {Tien}\ and\ \citenamefont {Koump}(1970)}]{Tien1970}%
  \BibitemOpen
  \bibfield  {author} {\bibinfo {author} {\bibfnamefont {R.~H.}\ \bibnamefont
  {Tien}}\ and\ \bibinfo {author} {\bibfnamefont {V.}~\bibnamefont {Koump}},\
  }\href {\doibase 10.1115/1.3449601} {\bibfield  {journal} {\bibinfo
  {journal} {Journal of Heat Transfer}\ }\textbf {\bibinfo {volume} {92}},\
  \bibinfo {pages} {11} (\bibinfo {year} {1970})}\BibitemShut {NoStop}%
\bibitem [{\citenamefont {Oxtoby}\ and\ \citenamefont
  {Harrowell}(1992)}]{doi:10.1063/1.462864}%
  \BibitemOpen
  \bibfield  {author} {\bibinfo {author} {\bibfnamefont {D.~W.}\ \bibnamefont
  {Oxtoby}}\ and\ \bibinfo {author} {\bibfnamefont {P.~R.}\ \bibnamefont
  {Harrowell}},\ }\href {\doibase 10.1063/1.462864} {\bibfield  {journal}
  {\bibinfo  {journal} {J. Chem. Phys.}\ }\textbf {\bibinfo {volume} {96}},\
  \bibinfo {pages} {3834} (\bibinfo {year} {1992})},\ \Eprint
  {http://arxiv.org/abs/https://doi.org/10.1063/1.462864}
  {https://doi.org/10.1063/1.462864} \BibitemShut {NoStop}%
\bibitem [{\citenamefont {Conti}(2001)}]{PhysRevE.64.051601}%
  \BibitemOpen
  \bibfield  {author} {\bibinfo {author} {\bibfnamefont {M.}~\bibnamefont
  {Conti}},\ }\href {\doibase 10.1103/PhysRevE.64.051601} {\bibfield  {journal}
  {\bibinfo  {journal} {Phys. Rev. E}\ }\textbf {\bibinfo {volume} {64}},\
  \bibinfo {pages} {051601} (\bibinfo {year} {2001})}\BibitemShut {NoStop}%
\bibitem [{\citenamefont {Conti}\ and\ \citenamefont
  {Fermani}(2003)}]{PhysRevE.67.026117}%
  \BibitemOpen
  \bibfield  {author} {\bibinfo {author} {\bibfnamefont {M.}~\bibnamefont
  {Conti}}\ and\ \bibinfo {author} {\bibfnamefont {M.}~\bibnamefont
  {Fermani}},\ }\href {\doibase 10.1103/PhysRevE.67.026117} {\bibfield
  {journal} {\bibinfo  {journal} {Phys. Rev. E}\ }\textbf {\bibinfo {volume}
  {67}},\ \bibinfo {pages} {026117} (\bibinfo {year} {2003})}\BibitemShut
  {NoStop}%
\bibitem [{\citenamefont {Conti}(2004)}]{PhysRevE.69.022601}%
  \BibitemOpen
  \bibfield  {author} {\bibinfo {author} {\bibfnamefont {M.}~\bibnamefont
  {Conti}},\ }\href {\doibase 10.1103/PhysRevE.69.022601} {\bibfield  {journal}
  {\bibinfo  {journal} {Phys. Rev. E}\ }\textbf {\bibinfo {volume} {69}},\
  \bibinfo {pages} {022601} (\bibinfo {year} {2004})}\BibitemShut {NoStop}%
\bibitem [{\citenamefont {Lowengrub}\ and\ \citenamefont
  {Truskinovsky}(1998)}]{Lowengrub2617}%
  \BibitemOpen
  \bibfield  {author} {\bibinfo {author} {\bibfnamefont {J.}~\bibnamefont
  {Lowengrub}}\ and\ \bibinfo {author} {\bibfnamefont {L.}~\bibnamefont
  {Truskinovsky}},\ }\href {\doibase 10.1098/rspa.1998.0273} {\bibfield
  {journal} {\bibinfo  {journal} {Proc. Roy. Soc. London A}\ }\textbf {\bibinfo
  {volume} {454}},\ \bibinfo {pages} {2617} (\bibinfo {year}
  {1998})}\BibitemShut {NoStop}%
\bibitem [{\citenamefont {Steinhardt}\ \emph {et~al.}(1983)\citenamefont
  {Steinhardt}, \citenamefont {Nelson},\ and\ \citenamefont
  {Ronchetti}}]{PhysRevB.28.784}%
  \BibitemOpen
  \bibfield  {author} {\bibinfo {author} {\bibfnamefont {P.~J.}\ \bibnamefont
  {Steinhardt}}, \bibinfo {author} {\bibfnamefont {D.~R.}\ \bibnamefont
  {Nelson}}, \ and\ \bibinfo {author} {\bibfnamefont {M.}~\bibnamefont
  {Ronchetti}},\ }\href {\doibase 10.1103/PhysRevB.28.784} {\bibfield
  {journal} {\bibinfo  {journal} {Phys. Rev. B}\ }\textbf {\bibinfo {volume}
  {28}},\ \bibinfo {pages} {784} (\bibinfo {year} {1983})}\BibitemShut
  {NoStop}%
\bibitem [{\citenamefont {Lechner}\ and\ \citenamefont
  {Dellago}(2008)}]{doi:10.1063/1.2977970}%
  \BibitemOpen
  \bibfield  {author} {\bibinfo {author} {\bibfnamefont {W.}~\bibnamefont
  {Lechner}}\ and\ \bibinfo {author} {\bibfnamefont {C.}~\bibnamefont
  {Dellago}},\ }\href {\doibase 10.1063/1.2977970} {\bibfield  {journal}
  {\bibinfo  {journal} {The Journal of Chemical Physics}\ }\textbf {\bibinfo
  {volume} {129}},\ \bibinfo {pages} {114707} (\bibinfo {year} {2008})},\
  \Eprint {http://arxiv.org/abs/https://doi.org/10.1063/1.2977970}
  {https://doi.org/10.1063/1.2977970} \BibitemShut {NoStop}%
\bibitem [{\citenamefont {Sasa}(2014)}]{PhysRevLett.112.100602}%
  \BibitemOpen
  \bibfield  {author} {\bibinfo {author} {\bibfnamefont {S.-i.}\ \bibnamefont
  {Sasa}},\ }\href {\doibase 10.1103/PhysRevLett.112.100602} {\bibfield
  {journal} {\bibinfo  {journal} {Phys. Rev. Lett.}\ }\textbf {\bibinfo
  {volume} {112}},\ \bibinfo {pages} {100602} (\bibinfo {year}
  {2014})}\BibitemShut {NoStop}%
\bibitem [{\citenamefont {Khinchin}\ and\ \citenamefont
  {Gamow}(1949)}]{KhinchinGamow}%
  \BibitemOpen
  \bibfield  {author} {\bibinfo {author} {\bibfnamefont {A.~I.}\ \bibnamefont
  {Khinchin}}\ and\ \bibinfo {author} {\bibfnamefont {G.}~\bibnamefont
  {Gamow}},\ }\href@noop {} {\emph {\bibinfo {title} {Mathematical Foundations
  of Statistical Mechanics}}}\ (\bibinfo  {publisher} {Dover Publications,
  Inc.},\ \bibinfo {address} {New York},\ \bibinfo {year} {1949})\BibitemShut
  {NoStop}%
\bibitem [{\citenamefont {Kikkinides}\ and\ \citenamefont
  {Monson}(2015)}]{doi:10.1063/1.4913636}%
  \BibitemOpen
  \bibfield  {author} {\bibinfo {author} {\bibfnamefont {E.~S.}\ \bibnamefont
  {Kikkinides}}\ and\ \bibinfo {author} {\bibfnamefont {P.~A.}\ \bibnamefont
  {Monson}},\ }\href {\doibase 10.1063/1.4913636} {\bibfield  {journal}
  {\bibinfo  {journal} {The Journal of Chemical Physics}\ }\textbf {\bibinfo
  {volume} {142}},\ \bibinfo {pages} {094706} (\bibinfo {year} {2015})},\
  \Eprint {http://arxiv.org/abs/https://doi.org/10.1063/1.4913636}
  {https://doi.org/10.1063/1.4913636} \BibitemShut {NoStop}%
\bibitem [{\citenamefont {Yang}\ \emph {et~al.}(1976)\citenamefont {Yang},
  \citenamefont {Fleming},\ and\ \citenamefont {Gibbs}}]{doi:10.1063/1.432687}%
  \BibitemOpen
  \bibfield  {author} {\bibinfo {author} {\bibfnamefont {A.~J.~M.}\
  \bibnamefont {Yang}}, \bibinfo {author} {\bibfnamefont {P.~D.}\ \bibnamefont
  {Fleming}}, \ and\ \bibinfo {author} {\bibfnamefont {J.~H.}\ \bibnamefont
  {Gibbs}},\ }\href {\doibase 10.1063/1.432687} {\bibfield  {journal} {\bibinfo
   {journal} {J. Chem. Phys.}\ }\textbf {\bibinfo {volume} {64}},\ \bibinfo
  {pages} {3732} (\bibinfo {year} {1976})},\ \Eprint
  {http://arxiv.org/abs/https://doi.org/10.1063/1.432687}
  {https://doi.org/10.1063/1.432687} \BibitemShut {NoStop}%
\bibitem [{\citenamefont {Goddard}\ \emph
  {et~al.}(2012{\natexlab{a}})\citenamefont {Goddard}, \citenamefont {Nold},
  \citenamefont {Savva}, \citenamefont {Pavliotis},\ and\ \citenamefont
  {Kalliadasis}}]{PhysRevLett.109.120603}%
  \BibitemOpen
  \bibfield  {author} {\bibinfo {author} {\bibfnamefont {B.~D.}\ \bibnamefont
  {Goddard}}, \bibinfo {author} {\bibfnamefont {A.}~\bibnamefont {Nold}},
  \bibinfo {author} {\bibfnamefont {N.}~\bibnamefont {Savva}}, \bibinfo
  {author} {\bibfnamefont {G.~A.}\ \bibnamefont {Pavliotis}}, \ and\ \bibinfo
  {author} {\bibfnamefont {S.}~\bibnamefont {Kalliadasis}},\ }\href {\doibase
  10.1103/PhysRevLett.109.120603} {\bibfield  {journal} {\bibinfo  {journal}
  {Phys. Rev. Lett.}\ }\textbf {\bibinfo {volume} {109}},\ \bibinfo {pages}
  {120603} (\bibinfo {year} {2012}{\natexlab{a}})}\BibitemShut {NoStop}%
\bibitem [{\citenamefont {Goddard}\ \emph
  {et~al.}(2012{\natexlab{b}})\citenamefont {Goddard}, \citenamefont {Nold},
  \citenamefont {Savva}, \citenamefont {Yatsyshin},\ and\ \citenamefont
  {Kalliadasis}}]{Goddard_2012}%
  \BibitemOpen
  \bibfield  {author} {\bibinfo {author} {\bibfnamefont {B.~D.}\ \bibnamefont
  {Goddard}}, \bibinfo {author} {\bibfnamefont {A.}~\bibnamefont {Nold}},
  \bibinfo {author} {\bibfnamefont {N.}~\bibnamefont {Savva}}, \bibinfo
  {author} {\bibfnamefont {P.}~\bibnamefont {Yatsyshin}}, \ and\ \bibinfo
  {author} {\bibfnamefont {S.}~\bibnamefont {Kalliadasis}},\ }\href {\doibase
  10.1088/0953-8984/25/3/035101} {\bibfield  {journal} {\bibinfo  {journal}
  {Journal of Physics: Condensed Matter}\ }\textbf {\bibinfo {volume} {25}},\
  \bibinfo {pages} {035101} (\bibinfo {year} {2012}{\natexlab{b}})}\BibitemShut
  {NoStop}%
\bibitem [{\citenamefont {Schmidt}\ and\ \citenamefont
  {Brader}(2013)}]{doi:10.1063/1.4807586}%
  \BibitemOpen
  \bibfield  {author} {\bibinfo {author} {\bibfnamefont {M.}~\bibnamefont
  {Schmidt}}\ and\ \bibinfo {author} {\bibfnamefont {J.~M.}\ \bibnamefont
  {Brader}},\ }\href {\doibase 10.1063/1.4807586} {\bibfield  {journal}
  {\bibinfo  {journal} {The Journal of Chemical Physics}\ }\textbf {\bibinfo
  {volume} {138}},\ \bibinfo {pages} {214101} (\bibinfo {year} {2013})},\
  \Eprint {http://arxiv.org/abs/https://doi.org/10.1063/1.4807586}
  {https://doi.org/10.1063/1.4807586} \BibitemShut {NoStop}%
\bibitem [{\citenamefont {Schmidt}(2018)}]{doi:10.1063/1.5008608}%
  \BibitemOpen
  \bibfield  {author} {\bibinfo {author} {\bibfnamefont {M.}~\bibnamefont
  {Schmidt}},\ }\href {\doibase 10.1063/1.5008608} {\bibfield  {journal}
  {\bibinfo  {journal} {The Journal of Chemical Physics}\ }\textbf {\bibinfo
  {volume} {148}},\ \bibinfo {pages} {044502} (\bibinfo {year} {2018})},\
  \Eprint {http://arxiv.org/abs/https://doi.org/10.1063/1.5008608}
  {https://doi.org/10.1063/1.5008608} \BibitemShut {NoStop}%
\bibitem [{\citenamefont {Korteweg}(1901)}]{Korteweg1901}%
  \BibitemOpen
  \bibfield  {author} {\bibinfo {author} {\bibfnamefont {D.~J.}\ \bibnamefont
  {Korteweg}},\ }\href@noop {} {\bibfield  {journal} {\bibinfo  {journal}
  {Arch. Neerl. Sci. Ex. Nat.}\ }\textbf {\bibinfo {volume} {6}},\ \bibinfo
  {pages} {1} (\bibinfo {year} {1901})}\BibitemShut {NoStop}%
\bibitem [{\citenamefont {Binder}(1973)}]{PhysRevB.8.3423}%
  \BibitemOpen
  \bibfield  {author} {\bibinfo {author} {\bibfnamefont {K.}~\bibnamefont
  {Binder}},\ }\href {\doibase 10.1103/PhysRevB.8.3423} {\bibfield  {journal}
  {\bibinfo  {journal} {Phys. Rev. B}\ }\textbf {\bibinfo {volume} {8}},\
  \bibinfo {pages} {3423} (\bibinfo {year} {1973})}\BibitemShut {NoStop}%
\bibitem [{\citenamefont {Wheeler}\ and\ \citenamefont
  {McFadden}(1997)}]{Wheeler08081997}%
  \BibitemOpen
  \bibfield  {author} {\bibinfo {author} {\bibfnamefont {A.~A.}\ \bibnamefont
  {Wheeler}}\ and\ \bibinfo {author} {\bibfnamefont {G.~B.}\ \bibnamefont
  {McFadden}},\ }\href {\doibase 10.1098/rspa.1997.0086} {\bibfield  {journal}
  {\bibinfo  {journal} {Proc. Roy. Soc. London A}\ }\textbf {\bibinfo {volume}
  {453}},\ \bibinfo {pages} {1611} (\bibinfo {year} {1997})}\BibitemShut
  {NoStop}%
\bibitem [{\citenamefont {Anderson}\ \emph {et~al.}(1998)\citenamefont
  {Anderson}, \citenamefont {McFadden},\ and\ \citenamefont
  {Wheeler}}]{doi:10.1146/annurev.fluid.30.1.139}%
  \BibitemOpen
  \bibfield  {author} {\bibinfo {author} {\bibfnamefont {D.~M.}\ \bibnamefont
  {Anderson}}, \bibinfo {author} {\bibfnamefont {G.~B.}\ \bibnamefont
  {McFadden}}, \ and\ \bibinfo {author} {\bibfnamefont {A.~A.}\ \bibnamefont
  {Wheeler}},\ }\href {\doibase 10.1146/annurev.fluid.30.1.139} {\bibfield
  {journal} {\bibinfo  {journal} {Annu. Rev. Fluid Mech.}\ }\textbf {\bibinfo
  {volume} {30}},\ \bibinfo {pages} {139} (\bibinfo {year} {1998})}\BibitemShut
  {NoStop}%
\bibitem [{\citenamefont {Anderson}\ \emph {et~al.}(2000)\citenamefont
  {Anderson}, \citenamefont {McFadden},\ and\ \citenamefont
  {Wheeler}}]{Anderson2000175}%
  \BibitemOpen
  \bibfield  {author} {\bibinfo {author} {\bibfnamefont {D.}~\bibnamefont
  {Anderson}}, \bibinfo {author} {\bibfnamefont {G.}~\bibnamefont {McFadden}},
  \ and\ \bibinfo {author} {\bibfnamefont {A.}~\bibnamefont {Wheeler}},\ }\href
  {\doibase http://dx.doi.org/10.1016/S0167-2789(99)00109-8} {\bibfield
  {journal} {\bibinfo  {journal} {Physica D}\ }\textbf {\bibinfo {volume}
  {135}},\ \bibinfo {pages} {175 } (\bibinfo {year} {2000})}\BibitemShut
  {NoStop}%
\bibitem [{\citenamefont {Langer}(1986)}]{Langer}%
  \BibitemOpen
  \bibfield  {author} {\bibinfo {author} {\bibfnamefont {J.~S.}\ \bibnamefont
  {Langer}},\ }\enquote {\bibinfo {title} {Models of pattern formation in
  first-order phase transitions},}\ in\ \href@noop {} {\emph {\bibinfo
  {booktitle} {Directions in Condensed Matter Physics}}}\ (\bibinfo {year}
  {1986})\ pp.\ \bibinfo {pages} {165--186}\BibitemShut {NoStop}%
\bibitem [{\citenamefont {Provatas}\ and\ \citenamefont
  {Elder}(2010)}]{Provatas:2010:PMM:1965378}%
  \BibitemOpen
  \bibfield  {author} {\bibinfo {author} {\bibfnamefont {N.}~\bibnamefont
  {Provatas}}\ and\ \bibinfo {author} {\bibfnamefont {K.}~\bibnamefont
  {Elder}},\ }\href@noop {} {\emph {\bibinfo {title} {Phase-Field Methods in
  Materials Science and Engineering}}},\ \bibinfo {edition} {1st}\ ed.\
  (\bibinfo  {publisher} {Wiley-VCH},\ \bibinfo {year} {2010})\BibitemShut
  {NoStop}%
\bibitem [{\citenamefont {Wilson}(1900)}]{doi:10.1080/14786440009463908}%
  \BibitemOpen
  \bibfield  {author} {\bibinfo {author} {\bibfnamefont {H.~W.}\ \bibnamefont
  {Wilson}},\ }\href {\doibase 10.1080/14786440009463908} {\bibfield  {journal}
  {\bibinfo  {journal} {Philos. Mag.}\ }\textbf {\bibinfo {volume} {50}},\
  \bibinfo {pages} {238} (\bibinfo {year} {1900})},\ \Eprint
  {http://arxiv.org/abs/https://doi.org/10.1080/14786440009463908}
  {https://doi.org/10.1080/14786440009463908} \BibitemShut {NoStop}%
\bibitem [{\citenamefont {Frenkel}(1932)}]{Frenkel}%
  \BibitemOpen
  \bibfield  {author} {\bibinfo {author} {\bibfnamefont {Y.~I.}\ \bibnamefont
  {Frenkel}},\ }\href@noop {} {\bibfield  {journal} {\bibinfo  {journal} {Phys.
  Z. Sowjetunion}\ }\textbf {\bibinfo {volume} {1}},\ \bibinfo {pages} {498}
  (\bibinfo {year} {1932})}\BibitemShut {NoStop}%
\bibitem [{\citenamefont {Kim}\ and\ \citenamefont
  {Lowengrub}(2005)}]{KimLowengrub2005}%
  \BibitemOpen
  \bibfield  {author} {\bibinfo {author} {\bibfnamefont {J.}~\bibnamefont
  {Kim}}\ and\ \bibinfo {author} {\bibfnamefont {J.}~\bibnamefont
  {Lowengrub}},\ }\href {\doibase 10.4171/IFB/132} {\bibfield  {journal}
  {\bibinfo  {journal} {Intf. Free Bound.}\ }\textbf {\bibinfo {volume} {7}},\
  \bibinfo {pages} {435 } (\bibinfo {year} {2005})}\BibitemShut {NoStop}%
\bibitem [{\citenamefont {Zaccarelli}\ \emph {et~al.}(2002)\citenamefont
  {Zaccarelli}, \citenamefont {Foffi}, \citenamefont {Gregorio}, \citenamefont
  {Sciortino}, \citenamefont {Tartaglia},\ and\ \citenamefont
  {Dawson}}]{Zaccarelli_2002}%
  \BibitemOpen
  \bibfield  {author} {\bibinfo {author} {\bibfnamefont {E.}~\bibnamefont
  {Zaccarelli}}, \bibinfo {author} {\bibfnamefont {G.}~\bibnamefont {Foffi}},
  \bibinfo {author} {\bibfnamefont {P.~D.}\ \bibnamefont {Gregorio}}, \bibinfo
  {author} {\bibfnamefont {F.}~\bibnamefont {Sciortino}}, \bibinfo {author}
  {\bibfnamefont {P.}~\bibnamefont {Tartaglia}}, \ and\ \bibinfo {author}
  {\bibfnamefont {K.~A.}\ \bibnamefont {Dawson}},\ }\href {\doibase
  10.1088/0953-8984/14/9/330} {\bibfield  {journal} {\bibinfo  {journal}
  {Journal of Physics: Condensed Matter}\ }\textbf {\bibinfo {volume} {14}},\
  \bibinfo {pages} {2413} (\bibinfo {year} {2002})}\BibitemShut {NoStop}%
\bibitem [{\citenamefont {T\'oth}\ \emph {et~al.}(2015)\citenamefont {T\'oth},
  \citenamefont {Pusztai},\ and\ \citenamefont
  {Gr\'an\'asy}}]{PhysRevB.92.184105}%
  \BibitemOpen
  \bibfield  {author} {\bibinfo {author} {\bibfnamefont {G.~I.}\ \bibnamefont
  {T\'oth}}, \bibinfo {author} {\bibfnamefont {T.}~\bibnamefont {Pusztai}}, \
  and\ \bibinfo {author} {\bibfnamefont {L.}~\bibnamefont {Gr\'an\'asy}},\
  }\href {\doibase 10.1103/PhysRevB.92.184105} {\bibfield  {journal} {\bibinfo
  {journal} {Phys. Rev. B}\ }\textbf {\bibinfo {volume} {92}},\ \bibinfo
  {pages} {184105} (\bibinfo {year} {2015})}\BibitemShut {NoStop}%
\end{thebibliography}%

\end{document}